\documentclass[11p,a4paper]{article}
\pdfoutput=1
\usepackage{jheppub}
\usepackage[utf8]{inputenc}
\usepackage{latexsym,amsfonts,amsmath,amscd,amssymb,epsf}
\usepackage{hyperref}
\usepackage[dvipsnames]{xcolor}
\usepackage{soul}
\usepackage{natbib}
\usepackage{graphicx}

\newcommand{\beq}{\begin{equation}}
\newcommand{\eeq}{\end{equation}}

\newcommand{\bear}{\begin{eqnarray}}
\newcommand{\eear}{\end{eqnarray}}

\newcommand{\cP}{{\cal P}}

\newcommand{\cK}{{\cal K}}

\usepackage{bm}
\let\vec=\bm

\renewcommand\Re{\operatorname{Re}}

\title{Medium-induced gluon radiation with full resummation of multiple scatterings for realistic parton-medium interactions}

\author[a]{Carlota Andres,}
\author[b,c]{Liliana Apolin\'ario,}
\author[d]{and Fabio Dominguez}
\emailAdd{carlota@jlab.org}
\emailAdd{liliana@lip.pt}
\emailAdd{fabio.dominguez@usc.es}

\preprint{JLAB-THY-20-3143}

\affiliation[a]{Jefferson Lab, Newport News, Virginia 23606, US}
\affiliation[b]{LIP, Av. Prof. Gama Pinto, 2, P-1649-003 Lisboa, Portugal}
\affiliation[c]{Instituto Superior T\'{e}cnico (IST), Universidade de Lisboa, Avenida Rovisco Pais 1, 1049-001 Lisbon, Portugal}
\affiliation[d]{Instituto Galego de F\'isica de Altas Enerx\'ias IGFAE, Universidade de Santiago de Compostela, E-15782 Santiago de Compostela (Galicia-Spain)}
\date{January 2020}

\begin{document}

\abstract{
The new precision era of jet quenching observables at both RHIC and the LHC calls for an improved and more precise description of in-medium gluon emissions. The development of new theoretical tools and analytical calculations to tackle this challenge has been hampered by the inability to include the effects of multiple scatterings with the medium using a realistic model for the parton-medium interactions. In this paper, we show how the analytical expressions for the full in-medium spectrum, including the resummation of all multiple scatterings, can be written in a form where the numerical evaluation can be easily performed without the need of the usually employed harmonic or single hard approximations. We present the transverse momentum and energy-dependent medium-induced gluon emission distributions for known realistic interaction models to illustrate how our framework can be applied beyond the limited kinematic regions of previous calculations.}

\maketitle

\section{Introduction}

The analysis of hard probes observables in high-energy nucleus-nucleus collisions has been proven to be one of the main tools at our disposal for the understanding and characterization of the properties of the quark-gluon plasma formed in these collisions. Recently, it has been recognized that besides the familiar studies of jet and hadron suppression, the theory of in-medium gluon emissions also plays a central role in the accurate description of a varied number of observables which can provide valuable information on the different stages of the evolution of the plasma created after the collision \cite{Apolinario:2017sob,Andres:2019eus}.

Given the importance of medium-induced radiation in the description of heavy-ion collisions, it is necessary to revisit and improve the current implementations of the evaluation of the gluon emission spectrum in a hot plasma. The level of accuracy achieved by the experimental data needs to be matched by the theoretical calculations. Therefore it is of the utmost importance to relax approximations and include all the physical effects in the formalism under which the in-medium radiation spectrum is computed.

The main difficulty encountered when an analytic approach is employed to calculate the emission spectrum off a hard parton is the appropriate inclusion of an arbitrary number of scatterings with the medium. It has long been known that multiple scatterings act coherently over high-energy particles modifying the emission spectrum through the largely studied Landau-Pomeranchuk-Migdal (LPM) effect \cite{Landau:1953um,Migdal:1956tc}. Going beyond the single particle spectrum, it has been shown that properly accounting for the effects of multiple scatterings is crucial for the description of processes with several particles, where color correlations are important and can be broken through interactions with the medium \cite{MehtarTani:2010ma,MehtarTani:2011tz,MehtarTani:2012cy}.

The usual approach to include the effect of multiple scatterings is to use the BDMPS-Z formalism \cite{Baier:1996kr, Baier:1996sk,Zakharov:1996fv,Zakharov:1997uu}, where a formal resummation can be achieved by considering only the non-relativistic dynamics in the transverse plane. When an expansion in terms of the number of scatterings is considered, this formalism has been shown to reproduce the outcomes obtained by different methods in which only a few scatterings are computed \cite{Wiedemann:2000za}. The problem with such approach lies in the fact that the formal expressions for the all-order resummation can be analytically computed only under very specific approximations which may miss some important physical effects. This is the case of the \emph{harmonic} or \emph{multiple soft scattering} approximation, which assumes a Gaussian profile for the transverse momentum transfers and does not reproduce the perturbative tails at high transverse momentum $k_\perp$. Given the impossibility of having fully analytical expressions, numerical implementations would be highly desirable, but so far, the only thorough attempts to numerically evaluate the emission spectrum and its $k_\perp$-dependence have been through computationally costly Monte Carlo implementations \cite{Feal:2018sml}.

In this paper, following the direction of ref.~\cite{CaronHuot:2010bp}, we show how to numerically evaluate the formal expressions of the medium-induced gluon emission spectrum for an arbitrary number of scatterings and realistic parton-medium interactions, without any further approximations. In order to do so, we numerically solve the appropriate differential equations defining the in-medium propagators which enter in the spectrum expression. In this way, we are able to successfully consider the full transverse momentum dependence and calculate the full emission spectrum, accounting for the relevant kinematic constraints.

The manuscript is organized as follows: in section~\ref{sec:BDMPS} we outline the fundamental assumptions of the BDMPS-Z approach as well as the limitations of its current approximate analytical evaluations. In section~\ref{sec:Setup} we present our framework, which allows us to calculate exactly the in-medium spectrum including  the  resummation  of  all  multiple  scatterings  for realistic parton-medium interactions. Readers not interested in all the details of the derivation can go directly to section~\ref{subsec:diffeqs} and section~\ref{subsec:energyspectrum}, which contain all the equations needed to compute the full resummed medium-induced $k_{\perp}$-differential and energy spectra. In section~\ref{sec:results} we present the numerical results  of  our  approach for two models of parton-medium interaction: Yukawa-type and hard thermal loop (HTL) interaction. Finally, we summarize and conclude in section~\ref{sec:conclusions}.

\section{Medium-induced gluon spectrum}
\label{sec:BDMPS}

The medium-induced gluon radiation spectrum in the high-energy limit has been derived in several formalisms \cite{Baier:1996kr,Zakharov:1996fv,Gyulassy:2000er,Arnold:2001ms}. For clarity, we will summarize the common assumptions entering these derivations and cite the result in the BDMPS-Z framework. Details about the derivations can be found in \cite{Baier:1996kr,Baier:1996sk,Zakharov:1996fv,Zakharov:1997uu,Wiedemann:2000za,CasalderreySolana:2007pr}. The basic assumptions which play an important role in the successful resummation of multiple scatterings into a compact formula are as follows:
\begin{itemize}
\item The opening angle of the radiation is small and the emission vertices are given by leading-order DGLAP splitting functions. Parent and daughter partons can pick up some momentum transverse to the direction of propagation of the initial parton, but their magnitudes are always much smaller than their respective energies.
\item The main contribution to radiation comes from elastic scatterings. The interaction between parton and medium is mediated by soft gluons, which are regarded as carrying only transverse momenta of the order of the characteristic medium scale.
\item At high-energy, the time scale of any single interaction is much shorter than the formation time of emitted gluons or the time scale for medium evolution. Therefore, the interactions are considered as instantaneous and are calculated for a fixed, but arbitrary, medium configuration which will be averaged over.
\end{itemize}

The specific details of the parton-medium interaction are given as a phenomenological input through the elastic collision rate $V(\vec{q})$, which then enters the calculation through the dipole cross section\footnote{Throughout, bold symbols describe two-dimensional variables and we adopt the shorthand $\int_{\vec{p}}= \int d^2\vec{p}/(2\pi)^2$ for the transverse integrals in momentum space.}
\beq
\sigma(\vec{r}) = \int_{\vec{q}} V(\vec{q}) \left(1 - e^{i \vec{q}\vec{r}}\right) \: .
\label{eq:sigmaV_coordinate}
\eeq
No further assumptions are made on the form of $V(\vec{q})$, although in all realistic models it must have the power behavior $V(\vec{q}) \sim 1/\vec{q}^4$, which is a direct consequence of having point-like interactions with a Coulomb potential at short distances.

For simplicity, we assume the emitted gluon is soft with $\omega/E\ll 1$ where $\omega$ is the energy of the emitted gluon and $E$ the energy of the initial parton. This is not a general assumption for the derivation of the formula for the spectrum and will be relaxed in a subsequent publication.\footnote{Details on how to properly incorporate into the differential spectrum the case when the gluon takes a finite energy fraction can found in \cite{Blaizot:2012fh,Apolinario:2014csa,Apolinario:2012vy}.} In this limit, the medium-induced gluon spectrum off a high-energy parton reads:
\begin{align}
\omega\frac{dI}{d\omega d^2\vec{k}} &= \frac{2\alpha_s C_R}{(2\pi)^2\omega^2} \Re \int_0^\infty dt' \, \int_0^{t'} dt \,\int_{\vec{p}\vec{q}} \, 
 \vec{p} \cdot \vec{q} \,\,\widetilde{\cK}(t',\vec{q};t,\vec{p})  \cP(\infty,\vec{k};t',\vec{q})\:,
 \label{eq:bdmps}
\end{align}
where $\vec{k}$ is the two-dimensional transverse momentum of the emitted gluon. The variables $t$ and $t'$ correspond to the emission times\footnote{We refer to ``time'' as being the longitudinal coordinate along the medium.} in the amplitude and conjugate amplitude, respectively, $\widetilde\cK (t',\vec{q};t,\vec{p})$ is the emission kernel in momentum space, and $\cP(\infty,\vec{k};t',\vec{q})$ is the momentum broadening factor. The radiation off hard quarks or gluons differs by the Casimir factor $C_R = C_F = (N_c^2 -1)/2N_c$ or $C_R=C_A=N_c$, respectively. 

The Green’s function $\widetilde\cK (t',\vec{q};t,\vec{p})$ can be explicitly written in coordinate space as the following path integral
\begin{align}
\cK \left(t',\vec{z};t,\vec{y}\right) &\equiv \int_{\vec{p}\vec{q}} \,  e^{i\left(\vec{q}\cdot\vec{z}-\vec{p}\cdot\vec{y}\right)}\,\widetilde{\cK} \left(t',\vec{q};t,\vec{p}\right)\:\:  \nonumber \\
&= \int_{\vec{r}(t)=\vec{y}}^{\vec{r}(t')=\vec{z}}{\cal D}\vec{r}\exp\left[\int_t^{t'}ds\;\left(\frac{i\omega}{2}\dot{\vec{r}}^2-\frac{1}{2}n(s)\sigma(\vec{r})\right)\right]\:,
\label{eq:kernelexplicit}
\end{align}
while the momentum broadening factor is given by
\beq
\cP(t'',\vec{k};t',\vec{q}) \equiv \int d^2\vec{z} \, e^{-i \left( \vec{k}-\vec{q} \right)\cdot \vec{z}} \,
\exp \left\{-\frac{1}{2}\,\int_{t'}^{t''} \,ds \, n(s)\, \sigma(\vec{z}) \right\}\:,
\label{eq:broadexplicit}
\eeq
with $n(s)$ the linear medium density.

The numerical evaluation of the path integral in  eq.~(\ref{eq:kernelexplicit}) including all the multiple scatterings for a realistic collision rate $V(\vec{q})$ --- such as a Yukawa-like interaction --- has always posed technical problems, which could not be overcome until very recently with advanced Monte Carlo techniques \cite{Feal:2018sml}. For this reason, the spectrum in eq.~(\ref{eq:bdmps}) has historically been treated in two approximations in which an analytical expression for the kernel is possible: multiple soft and single hard momentum transfer.

Within a multiple soft in-medium scatterings approach, the dipole cross section can be approximated by its leading logarithmic behavior \beq
n(s)\sigma(\vec{r}) \approx \frac{1}{2} \hat{q}(s)\vec{r}^2  + \mathcal{O} (\vec{r}^2 \ln\vec{r}^2) \:,
\label{eq:qhat}
\eeq
where $\hat{q}$ is the transport coefficient that characterizes the average transverse momentum squared transferred from the medium to the projectile per unit path length. This approximation is  usually employed for opaque media, when configurations where the transverse distance $\vec{r}$ is large are believed to be strongly suppressed. By replacing eq.~(\ref{eq:qhat}) in eq.~(\ref{eq:kernelexplicit}), an analytical solution for the path integral is straightforward to obtain in the static case.\footnote{For expanding media the static solution is also of great use, since scaling relations have been shown to work for phenomenological purposes \cite{Baier:1996sk,Salgado:2002cd}.} This result is also known as the \emph{harmonic oscillator} (HO) or \emph{Gaussian} approximation. It is worth emphasizing that even though this approximation allows us to have analytical calculations which can provide insight and qualitative descriptions, it has not been proven to correspond to any limit in terms of physical variables, and therefore its quantitative predictions have to be taken with care. One of the major drawbacks of this method is the strong suppression of the high transverse momentum part of the spectrum where it has an exponential behavior instead of the power-like tails characteristic of Coulomb interactions at short distances.

The other scenario corresponds to the radiation pattern resulting from an incoherent superposition of just a few single hard scattering processes. This limit can be obtained by expanding the integrand of eq.~(\ref{eq:kernelexplicit}) in powers of the density of scattering centers $\left(n(s)\sigma(\vec{r})\right)^N$ \cite{Wiedemann:2000za,Gyulassy:2000er}. This approach is usually known as \emph{opacity expansion}. The first order ($N=1$) in this procedure is typically referred to as the Gyulassy-Levai-Vitev (GLV) or first opacity approximation and is applicable for dilute media. This approximation is expected to be suitable for gluons with transverse momenta larger than the characteristic scale of the medium and also for high-energy gluons. When the number of scattering centers is large, resuming the contributions from all orders in opacity is needed, a process that is both analytically and computationally demanding. 

The above mentioned differences between these two approaches have a direct essential consequence: the energy spectrum produced by the Gaussian approximation is much softer than the one produced by the opacity expansion. This is mainly due to two factors:  (1) the absence of destructive LPM interferences in the latter, which naturally appear in the former; and (2) the inclusion of the power-law tails of the interaction cross-section in the first opacity result that otherwise are neglected in the harmonic approximation.\footnote{A more thorough comparison of these two approximations can be found in \cite{Arnold:2009mr,Salgado:2003gb}.} The use of these approximate solutions have led to conflicting results when extracting medium properties from measurements taken at RHIC and at the LHC \cite{Andres:2016iys,Burke:2013yra}, while recent studies may indicate that when these approximations are not employed this centrality/energy puzzle seems to disappear \cite{Feal:2019xfl}. 

In this manuscript, we thus attempt to provide a framework that naturally includes and goes beyond both approaches above by avoiding any assumption on the nature of the interaction between the parton and the medium. In the following section, we will explain the logical setup that allows us to derive an analytical expression for the medium-induced gluon radiation in the most general case.

\section{Setting up the evaluation}
\label{sec:Setup}

Apart from the difficulties that arise when attempting to compute the kernel given by eq.~(\ref{eq:kernelexplicit}) without any further approximations, there are several obstacles in numerically evaluating the spectrum in eq.~(\ref{eq:bdmps}). Hence, it is convenient to rearrange eq.~(\ref{eq:bdmps}) to put it in a more appropriate form. 

One issue is that the gluon emission can occur anywhere after the initial parton is created, that is, either inside or outside the medium. In consequence, the interferences between these two types of emissions need to be properly taken into account. The common way of addressing this is by splitting the semi-infinite integration in the emission times into two terms bound by the length of the medium, which correspond to the pure in-medium emissions (usually denoted as ``in-in'' contributions) and the medium-vacuum interference (usually named ``in-out''). While this is not a limiting factor, it demands precise cancellations between these two types of contributions, which might involve an additional level of precision in the numerical evaluation, thus making it inefficient. To overcome this issue, we analytically perform the integration in $t'$ in eq.~(\ref{eq:bdmps}). This introduces in turn an integration over the position of one of the scatterings. Since no scatterings occur outside the medium, the resulting time integrations are naturally bound by its length and thus, separating them into ``in-in" and ``in-out" pieces is no longer necessary.

On the other hand, as it was explained in the previous section, the expressions of the broadening factor $\cP$ and the kernel $\widetilde\cK$ are naturally written in coordinate space, the latter involving a complicated path integral. To avoid the difficulties that arise when attempting to numerically compute this path integral, we work directly in momentum space by considering these objects as propagators that satisfy specific differential equations which can be numerically solved by conventional methods.

\subsection{Reorganization of the spectrum}
\label{subsec:reorganization}
We start by performing the $t'$ integration in the in-medium spectrum given by eq.~(\ref{eq:bdmps}).  Even though $t'$ is an argument of both $\widetilde\cK$ and $\cP$ in eq.~(\ref{eq:bdmps}), it can be integrated out without knowing the explicit form of either of these two factors. For that end, we only need to notice that both $\widetilde\cK$ and $\cP$ are propagators that satisfy the following Schwinger-Dyson type equations:
\begin{align}
\cP(t'',\vec{k};t',\vec{q}) &= \left(2\pi\right)^2\delta^{(2)}(\vec{k}-\vec{q})\,-\,\frac{1}{2}  \int_{t`}^{t''} ds \,n(s) \int_{\vec{k}'} \, \sigma(\vec{k}'-\vec{q}) \cP(t'',\vec{k};s,\vec{k}')\: ,\label{eq:PSD1}\\
\widetilde{\cK}(t',\vec{q};t,\vec{p}) &= \left(2\pi\right)^2\delta^{(2)}(\vec{q}-\vec{p})\,e^{-i \frac{p^2}{2\omega}(t'-t)}\nonumber\\
&\quad \qquad-\,\frac{1}{2}  \int_{t}^{t'} ds \,n(s) \int_{\vec{k}'}  \,\sigma(\vec{q}-\vec{k}') e^{-i \frac{q^2}{2\omega}(t'-s)} \widetilde{\cK}(s,\vec{k}';t,\vec{p})\:.
\label{eq:KSD1}
\end{align}

Here, the dependence in $t'$ is confined to phase factors and limits of integration. Now we can replace eqs.~(\ref{eq:PSD1})~and~(\ref{eq:KSD1}) in eq.~(\ref{eq:bdmps}) and perform the $t'$-integration analytically and, after some manipulations (shown in appendix~\ref{sec:derivation}), arrive at
\begin{align}
\omega\frac{dI}{d\omega d^2\vec{k}} = \frac{2\alpha_s C_R}{(2\pi)^2\omega} \Re\int_0^\infty ds\; n(s) &\int_0^s dt\int_{\vec{p}\vec{q}\vec{l}}i\vec{p}\cdot\left(\frac{\vec{l}}{\vec{l}^2}-\frac{\vec{q}}{\vec{q}^2}\right)\sigma(\vec{l}-\vec{q})\widetilde\cK(s,\vec{q};t,\vec{p})\cP(\infty,\vec{k};s,\vec{l})
\:,
\label{eq:fullspect}
\end{align}
where the vacuum contribution has already been subtracted. In this expression there is still one time integral running up to infinity, but with the main difference that it represents the position of one of the scatterings and hence the integrand is zero outside of the medium. This allows us to use the length of the medium $L$ as the upper limit for this integral and for the end point of the momentum broadening as well. The effect of emissions outside of the medium has already been integrated out (or subtracted in the case of the purely vacuum emissions) and there is no need to deal with interferences separately or rely on precise cancellations between different terms.

This new expression for the in-medium spectrum has another advantage with regard to its numerical evaluation. The dipole cross section behaves, for any realistic parton-medium interaction,  as $\sigma(\vec{q})\sim1/\vec{q}^4$ at large $\vec{q}$, which guarantees the convergence of the integrals over $\vec{q}$ and $\vec{l}$, while also providing a convenient initial condition which can be evolved when $\widetilde\cK$ and $\cP$ are taken as propagators.

It is worth noticing that extracting the first order in opacity result from eq.~(\ref{eq:fullspect}) is straightforward. We only need to take the vacuum versions of $\cP$ and $\widetilde\cK$, which can be read off directly from the first term in the r.h.s. of eqs.~(\ref{eq:PSD1})~and~(\ref{eq:KSD1}). This is done in more detail in appendix \ref{sec:glv}. The obtained results represent the large energy/transverse momentum limit of eq.~(\ref{eq:fullspect}), and as such, quantitative comparisons are possible. On the contrary, the Harmonic Oscillator cannot be obtained as a  particular limit of the physical variables in eq.~(\ref{eq:fullspect}), yielding an approximate result, more suitable for opaque media, which makes the comparison with the full solution more difficult. As such, in section \ref{sec:results}, when comparing the full resummed spectrum with the HO result we focus only on qualitative features, while for the GLV case, we perform a more quantitative comparison.

\subsection{Recasting the evaluation of the spectrum as a first order linear differential equations problem}

In this section we will describe in detail how to evaluate the differential spectrum of eq.~(\ref{eq:fullspect}) using evolution equations instead of finding explicit expressions for the kernel $\widetilde\cK$ and the broadening factor $\cP$.

First, we recognize that $\cP$ satisfies the following differential equation:
\beq
\partial_\tau \cP(\tau,\vec{k};s,\vec{l}) = -\frac{1}{2}n(\tau)\int_{\vec{k}'}\, \sigma(\vec{k}-\vec{k}')\cP(\tau,\vec{k}';s,\vec{l})\:.
\eeq
with initial condition $\cP(s,\vec{k};s,\vec{l})=(2\pi)^2\delta^{(2)}(\vec{k}-\vec{l})$. Instead of trying to solve this equation for $\cP$, we define (based on eq.~\eqref{eq:fullspect})
\beq
\vec{\phi}(\tau,\vec{k};s,\vec{q}) = n(s)\int_{\vec{l}}\left(\frac{\vec{l}}{\vec{l}^2}-\frac{\vec{q}}{\vec{q}^2}\right)\sigma(\vec{l}-\vec{q})\cP(\tau,\vec{k};s,\vec{l})\:.
\eeq
It is clear that $\vec{\phi}(\tau,\vec{k};s,\vec{q})$ satisfies
\beq
\partial_\tau \vec{\phi}(\tau,\vec{k};s,\vec{q}) = -\frac{1}{2}n(\tau)\int_{\vec{k}'}\, \sigma(\vec{k}-\vec{k}')\vec{\phi}(\tau,\vec{k}';s,\vec{q})\:,
\label{eq:diffphi}
\eeq
with initial condition
\beq
\vec{\phi}(s,\vec{k};s,\vec{q}) = n(s)\left(\frac{\vec{k}}{\vec{k}^2}-\frac{\vec{q}}{\vec{q}^2}\right)\sigma(\vec{k}-\vec{q})\:.
\label{eq:initphi}
\eeq

Now, we perform a similar manipulation for the kernel $\widetilde \cK$. First, we recognize that it satisfies the following differential equation
\beq
\partial_{t}\widetilde{\cK}(s,\vec{q};t,\vec{p}) = \frac{i\vec{p}^2}{2\omega}\widetilde{\cK}(s,\vec{q};t,\vec{p})+\frac{1}{2}n(t)\int_{\vec{k}'}\sigma(\vec{k}'-\vec{p})\widetilde{\cK}(s,\vec{q};t,\vec{k}')\:,
\label{eq:diffKt}
\eeq
with initial condition $\widetilde\cK(s,\vec{q};s,\vec{p})=(2\pi)^2\delta^{(2)}(\vec{q}-\vec{p})$. Then, we define
\beq
\vec{\psi}(s,\vec{k};t,\vec{p}) = \int_{\vec{q}}\vec{\phi}(L,\vec{k};s,\vec{q})\,\widetilde\cK(s,\vec{q};t,\vec{p})\:,
\eeq
which also satisfies
\beq
\partial_{t}\vec{\psi}(s,\vec{k};t,\vec{p}) = \frac{i\vec{p}^2}{2\omega}\vec{\psi}(s,\vec{k};t,\vec{p})+\frac{1}{2}n(t)\int_{\vec{k}'}\sigma(\vec{k}'-\vec{p})\vec{\psi}(s,\vec{k};t,\vec{k}')\:,
\label{eq:diffpsi}
\eeq
with initial condition $\vec{\psi}(s,\vec{k};s,\vec{p})=\vec{\phi}(L,\vec{k};s,\vec{p})$. The first term in the right-hand side of eq.~(\ref{eq:diffpsi}) causes oscillations which guarantee the convergence of the $\vec{p}$-integrals, but it might become a problem when attempting to numerically solve the differential equation. It is then convenient to switch to the interaction picture, with
\beq
\vec{\psi}_I(s,\vec{k};t,\vec{p}) = e^{\frac{i\vec{p}^2}{2\omega}(s-t)}\vec{\psi}(s,\vec{k};t,\vec{p})\:,
\eeq
satisfying
\beq
\partial_{t}\vec{\psi}_I(s,\vec{k};t,\vec{p}) = \frac{1}{2}n(t)\int_{\vec{k}'}e^{\frac{i\vec{p}^2}{2\omega}(s-t)}\sigma(\vec{k}'-\vec{p})e^{-\frac{i\vec{k}^{\prime2}}{2\omega}(s-t)}\vec{\psi}_I(s,\vec{k};t,\vec{k}')\:,
\label{eq:diffpsiI}
\eeq
with initial condition
\beq
\vec{\psi}_I(s,\vec{k};s,\vec{p}) = \vec{\phi}(L,\vec{k};s,\vec{p})\:.
\label{eq:initpsiI}
\eeq

The full $\vec{k}$-dependent spectrum can then be written as
\beq
\omega\frac{dI}{d\omega d^2\vec{k}} = \frac{2\alpha_s C_R}{(2\pi)^2\omega} \Re\int_0^L ds \int_0^s dt\int_{\vec{p}}ie^{-\frac{i\vec{p}^2}{2\omega}(s-t)}\vec{p}\cdot\vec{\psi}_I(s,\vec{k};t,\vec{p})\:.
\label{eg:specpsiI}
\eeq

The procedure to evaluate the spectrum is then clear. First, we start by computing eq.~(\ref{eq:initphi}), then numerically solve eq.~(\ref{eq:diffphi}) to get the r.h.s. of eq.~(\ref{eq:initpsiI}), which is the starting point to numerically solve eq.~(\ref{eq:diffpsiI}). Once we have a solution for $\vec{\psi}_I$, it can be plugged into eq.~(\ref{eg:specpsiI}) and integrated numerically to obtain the spectrum. Before we can use this procedure for the numerical evaluation a few more manipulations are needed. Let us recall the form of the dipole cross section $\sigma$, in momentum space, in terms of the collision rate $V$,
\beq
\sigma(\vec{q}) = - V(\vec{q}) + (2\pi)^2\delta^{(2)}(\vec{q})\int_{\vec{l}}V(\vec{l})\; .
\label{eq:sigmaV}
\eeq
Then, the differential equation eq.~(\ref{eq:diffphi}) and its initial condition eq.~(\ref{eq:initphi}) take the form
\begin{align}
\partial_\tau \vec{\phi}(\tau,\vec{k};s,\vec{q}) &= -\frac{1}{2}n(\tau)\int_{\vec{k}'}\,V(\vec{k}-\vec{k}')\left[\vec{\phi}(\tau,\vec{k};s,\vec{q})-\vec{\phi}(\tau,\vec{k}';s,\vec{q})\right]\:,\\
\vec{\phi}(s,\vec{k};s,\vec{q}) &= n(s)\left(\frac{\vec{q}}{\vec{q}^2}-\frac{\vec{k}}{\vec{k}^2}\right)V(\vec{k}-\vec{q})\:,
\end{align}
while the differential equation for $\vec{\psi}_I$ given in eq.~(\ref{eq:diffpsiI}) is now
\beq
\partial_{t}\vec{\psi}_I(s,\vec{k};t,\vec{p}) = \frac{1}{2}n(t)\int_{\vec{k}'}\, V(\vec{k}'-\vec{p})\left[\vec{\psi}_I(s,\vec{k};t,\vec{p})-e^{-\frac{i(\vec{k}^{\prime2}-\vec{p}^2)}{2\omega}(s-t)}\vec{\psi}_I(s,\vec{k};t,\vec{k}')\right]\:.
\eeq

For most of the cases of interest the direction of $\vec{k}$ is irrelevant, thus, we can focus on the spectrum as a function only of its magnitude. We can therefore integrate over the direction of $\vec{k}$, which allows us to use rotational symmetry to analytically perform all angular integrals. The spectrum to evaluate is then
\beq
\omega\frac{dI}{d\omega dk^2} = \frac{1}{2}\int_0^{2\pi} d\theta_k\; \omega\frac{dI}{d\omega d^2\vec{k}}\:,
\eeq
which will be written in terms of the functions
\begin{align}
\label{eq:psiRescaled}
\frac{1}{2}\int_0^{2\pi}\frac{d\theta_k}{2\pi}\,\vec{\psi}_I(s,\vec{k};t,\vec{p}) &= \frac{\vec{p}}{\vec{p}^2}\tilde\psi_I(s,|\vec{k}|;t,|\vec{p}|)\; ,\\
\frac{1}{2}\int_0^{2\pi} \frac{d\theta_k}{2\pi}\; \vec{\phi}(\tau,\vec{k};s,\vec{q}) &= \frac{\vec{q}}{\vec{q}^2}\tilde\phi(\tau,|\vec{k}|;s,|\vec{q}|)\; .
\label{eq:phiRescaled}
\end{align}

\subsubsection{Set of equations to solve numerically}
\label{subsec:diffeqs}

For convenience, we change our variables to make them dimensionless: dummy momentum variables are rescaled as $\vec{p}\to\sqrt{2\omega/L}\,\vec{p}$ and time variables as $s\to Ls$. The typical transverse momentum transfer $\mu$, usually taken as the Debye mass of the screened interactions, sets the scale for the transverse momentum and the energy of the emitted gluons. The dimensionless variables in which we will evaluate the spectrum are\footnote{$\bar\omega_c$ is usually known as \emph{characteristic gluon frequency} and, as we will see later, the emission of gluons with $\omega > \bar \omega_c$ is suppressed.}
\beq
\kappa^2 = \frac{k^2}{\mu^2}\:, \qquad x=\frac{\omega}{\bar\omega_c}=\frac{2\omega}{\mu^2L}\:.
\label{eq:kappaAndx}
\eeq
Using the rescaled versions of $\tilde\psi_I$ and $\tilde\phi$ (see eqs.~\eqref{eq:psiRescaled}~and~\eqref{eq:phiRescaled}), we can define
\begin{align}
f_x(s,\kappa;t,p) &= \mu^2L\, \tilde\psi_I(sL,\mu\kappa;tL,p\sqrt{2\omega/L})\; , \\
g_x(\tau,l;s,q) &= 2\omega\,\tilde\phi(\tau L,l\sqrt{2\omega/L};sL,q\sqrt{2\omega/L})\; ,
\end{align}
and then write the differential medium-induced gluon emission spectrum as
\beq
x\frac{dI}{dx d\kappa^2} = \frac{\alpha_s C_R}{\pi^2}\Re\int_0^1 ds\int_0^s dt\int_0^\infty dp\;ip\,e^{-ip^2(s-t)} f_x(s,\kappa;t,p)\; ,
\label{eq:dimlessktspec}
\eeq
where $f_x$ satisfies the differential equation
\begin{align}
\partial_t f_x(s,\kappa;t,p) = \frac{1}{2}\tilde n(t)L\int_0^\infty&\frac{dq}{2\pi}\left[q\,\tilde V_1(q,p;x)f_x(s,\kappa;t,p) \right. \nonumber \\
&\left.- \,e^{-i(q^2-p^2)(s-t)}\,p\,\tilde V_2(q,p;x)f_x(s,\kappa;t,q)\right]\:,
\label{eq:dimlessktdiff}
\end{align}
with initial condition
\beq
f_x(s,\kappa;s,p) = \frac{1}{x}g_x(1,\kappa/\sqrt{x};s,p)\:,
\eeq
and $g_x$ satisfies the differential equation
\beq
\partial_\tau g_x(\tau,l;s,p) = -\frac{1}{2}\tilde n(\tau)L\int_0^\infty \frac{q\,dq}{2\pi} \, \tilde V_1(l,q;x)\left[g_x(\tau,l;s,p)-g_x(\tau,q;s,p)\right]\:,
\label{eq:diffeqg}
\eeq
with initial condition
\beq
g_x(s,l;s,p) = \frac{1}{2}\tilde n(s)L\left[\tilde V_1(l,p;x) - \frac{p}{l}\tilde V_2(l,p;x)\right]
\:.
\label{eq:initcondg}
\eeq

Here $\tilde n(s) = n(sL)$, while $\tilde V_1$ and $\tilde V_2$ are the two first angular moments of the rescaled collision rate
\begin{align}
\tilde V_1(|\vec{q}|,|\vec{p}|;x) &= \int_0^{2\pi}\frac{d\theta_{qp}}{2\pi}\;\tilde V(\vec{q}-\vec{p};x)\:,\label{eq:V1} \\
\tilde V_2(|\vec{q}|,|\vec{p}|;x) &= \int_0^{2\pi}\frac{d\theta_{qp}}{2\pi}\;\cos\theta_{qp}\,\tilde V(\vec{q}-\vec{p};x)\:,
\label{eq:V2}
\end{align}
where $\theta_{qp}$ is the angle between $\vec{q}$ and $\vec{p}$, and $\tilde V(\vec{q};x) = \frac{2\omega}{L}V(\vec{q}\sqrt{2\omega/L})$. For the usual models for the collision rate used in phenomenology, these angular integrals can be performed analytically, as will be shown in section~\ref{sec:results}.

\subsection{Energy spectrum}
\label{subsec:energyspectrum}

We can also integrate over transverse momentum to obtain the energy spectrum, always keeping in mind that the integration must respect the kinematical constraint $k \leq \omega$. We get
\begin{align}
x\frac{dI}{dx} &= \int_0^{\omega^2/\mu^2}d\kappa^2\;x\frac{dI}{dxd\kappa^2} \nonumber \\
&= \frac{\alpha_s C_R}{\pi^2}\Re\int_0^1 ds\int_0^s dt\int_0^\infty dp\;ip\,e^{-ip^2(s-t)} F_x(s,t;p)
\:.
\label{eq:speckinconst}
\end{align}
where
\beq
F_x(s,t;p) = \int_0^{\bar Rx^2/2} d\kappa^2\; f_x(s,\kappa;t,p)\:,
\eeq
and $\bar R = \bar\omega_cL = \frac{1}{2}\mu^2L^2$. It is clear that $F_x$ satisfies the differential equation eq.~(\ref{eq:dimlessktdiff}) with initial condition
\beq
F_x(s,s;p) = \int_0^{\bar Rx/2}dl^2\;g_x(1,l;s,p)
\:,
\label{eq:initcondfiniteR}
\eeq
while $g_x$ is still obtained by solving eq.~(\ref{eq:diffeqg}) with initial condition eq.~(\ref{eq:initcondg}). The case where the kinematical condition is removed, i.e. $\bar R\to\infty$ with fixed $\bar\omega_c$, is much simpler since the momentum broadening of the emitted gluon is irrelevant and therefore there is no need to solve eq.~(\ref{eq:diffeqg}). This can be seen directly from the equations by integrating $l$ from 0 to $\infty$ in eq.~(\ref{eq:diffeqg}) and noticing the right hand side vanishes. It is important to emphasize here that the derivation of the medium-induced emission spectrum (see section~\ref{sec:BDMPS}) assumes that the transverse momentum of the radiated gluon is small ($k\ll \omega$). Therefore, since the $\bar R\to\infty$ with fixed $\bar\omega_c$ limit is equivalent to extending the integration over the transverse momentum of the emitted gluon up to infinity, this limit does not correspond to any realistic physical situation and leads to a divergent spectrum for small values of $x$. Contrarily, when realistic kinematic constraints on the transverse momentum phase space ($\bar R$ finite) are imposed, the gluon energy distribution at small $x$ is depleted. In the following, we will present the results for $\bar R\to\infty$ along with the curves for finite values of $\bar R$ for illustrative purposes and as a check that our curves have the correct behavior at large values of $x$.

\section{Numerical results}
\label{sec:results}

In this section, we present the results of our numerical analysis. For simplicity, we perform our calculations in a static thermally equilibrated quark-gluon plasma and leave the extension to expanding media for subsequent publications. The linear density of scatterings is then a constant $n(t)=n_0 \Theta(L-t)$. For illustration purposes, we  always consider the case where the parent parton is a quark. Therefore we take $C_R=C_F=4/9$. The strong coupling is fixed to $\alpha_s=0.3$.

The numerical implementation of equations (\ref{eq:dimlessktspec})--(\ref{eq:initcondg}) involves momentum integrations that run up to infinity. The high-momentum tail of $V(\vec{q})\sim1/\vec{q}^4$ guarantees that all the integrands in eqs.~(\ref{eq:dimlessktspec})--(\ref{eq:initcondg}) approach zero as $1/\vec{q}^4$ (or faster), with increasing momenta. As such, the numerical evaluation uses an upper cut-off for these integrals, and we have carefully checked the stability of the result when this cut-off is changed.

In the following, we  study the results of our approach for two collision rate models. We first consider a Yukawa-type interaction and make straightforward comparisons with the respective first order in opacity. We also attempt to compare our results with the  Gaussian approximation, but always keeping in mind that there are subtle complications when attempting a direct correspondence between the parameters involved in both evaluations. Finally, we consider the case where the interaction is modeled through the collision rate calculated perturbatively in a hard thermal loop (HTL) formalism \cite{Aurenche:2002pd}.\footnote{The choice of models is motivated by the fact that they have the correct UV physics and are commonly used in phenomenological applications. Our approach is not restricted to this choice and one could in principle implement models with the correct IR limit as the lattice calculations in \cite{Moore:2019lgw}.}

\subsection{Yukawa-type interaction}
The collision rate $V$ for a Yukawa-type elastic scattering center is given by:
\beq
V(\vec{q}) = \frac{8\pi\mu^2}{(\vec{q}^2+\mu^2)^2}
\:,
\label{eq:yukawaPotential}
\eeq
where the screening mass $\mu$ is related to the Debye mass in a thermal medium, $\mu^2 \sim m_D^2$.

\begin{figure}[t]
\vspace*{-7mm}
\centering
\includegraphics[scale=0.37]{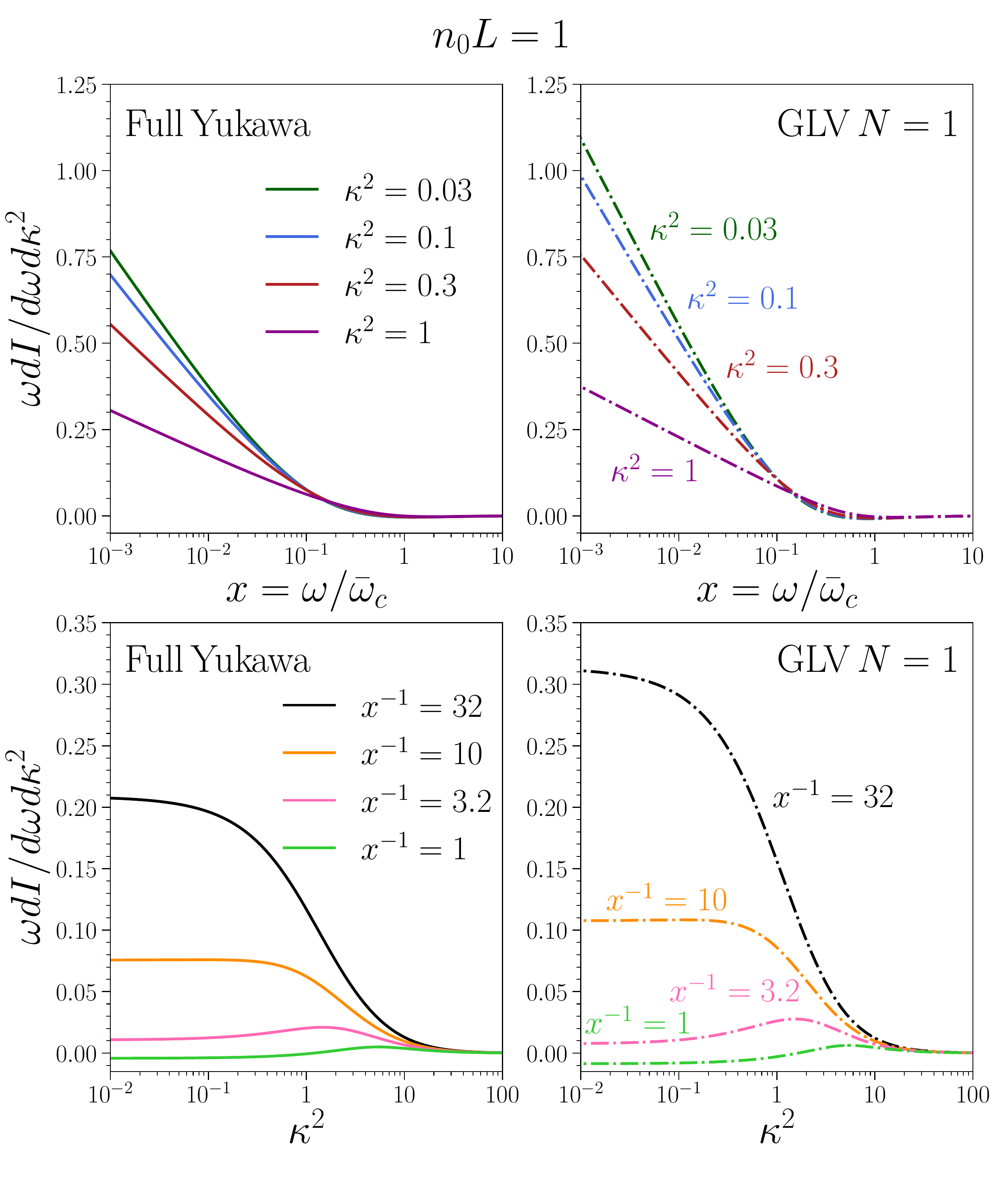}
\vspace*{-5mm}
\caption{Full medium-induced gluon radiation  $k{_\perp}$-differential spectrum for a medium with $n_0L=1$ for the Yukawa collision rate (left column) compared to the GLV first opacity approximation (right column). The upper panels show these spectra as function of the
rescaled gluon energy $x=\omega/\bar{\omega_c}$ for fixed values of the rescaled gluon transverse momentum $\kappa = k/\mu$. The lower panels show these spectra versus $\kappa^2$ for fixed values of
$x$. }
\label{fig:ktspec_yukawa_glv_n0L1}
\end{figure}

Its rescaled version $\tilde V$ is
\beq
\tilde V(\vec{q},x) = \frac{8\pi}{x(\vec{q}^2+1/x)^2}
\:,
\eeq
where $x = \frac{2\omega}{\mu^2L}$. 

Performing the angular integrations of eqs.~(\ref{eq:V1}) and (\ref{eq:V2}), yields
\begin{align}
\tilde V_1(q,p;x) &= \frac{4}{x}\int_0^{2\pi}\frac{d\theta}{(p^2+q^2-2pq\cos\theta+1/x)^2} \nonumber \\
\label{eq:integralAngle1}
&= \frac{8\pi(p^2+q^2+1/x)}{x[(p^2+q^2+1/x)^2-4p^2q^2]^{3/2}}\:, \\
\tilde V_2(q,p;x) &= \frac{4}{x}\int_0^{2\pi}\frac{d\theta \;\cos\theta}{(p^2+q^2-2pq\cos\theta+1/x)^2} \nonumber \\
\label{eq:integralAngle2}
&= \frac{16\pi pq}{x[(p^2+q^2+1/x)^2-4p^2q^2]^{3/2}}
\:.
\end{align}

Notice also that
\beq
\int_0^{\infty}\frac{dq}{2\pi}\,q\,\tilde V_1(q,p) = \int_{\vec{q}}\tilde V(\vec{q}-\vec{p}) = 2
\:.
\eeq

In fact, $V$ is normalized in this way to ensure that  $n_0L$ is the correct parameter for an opacity expansion.\footnote{Note that in the relevant equations eqs.~(\ref{eq:dimlessktspec})-(\ref{eq:initcondg}), $V$ always appears preceded by a factor $1/2$.}

\begin{figure}[t]
\vspace*{-5mm}
\centering
\includegraphics[scale=0.37]{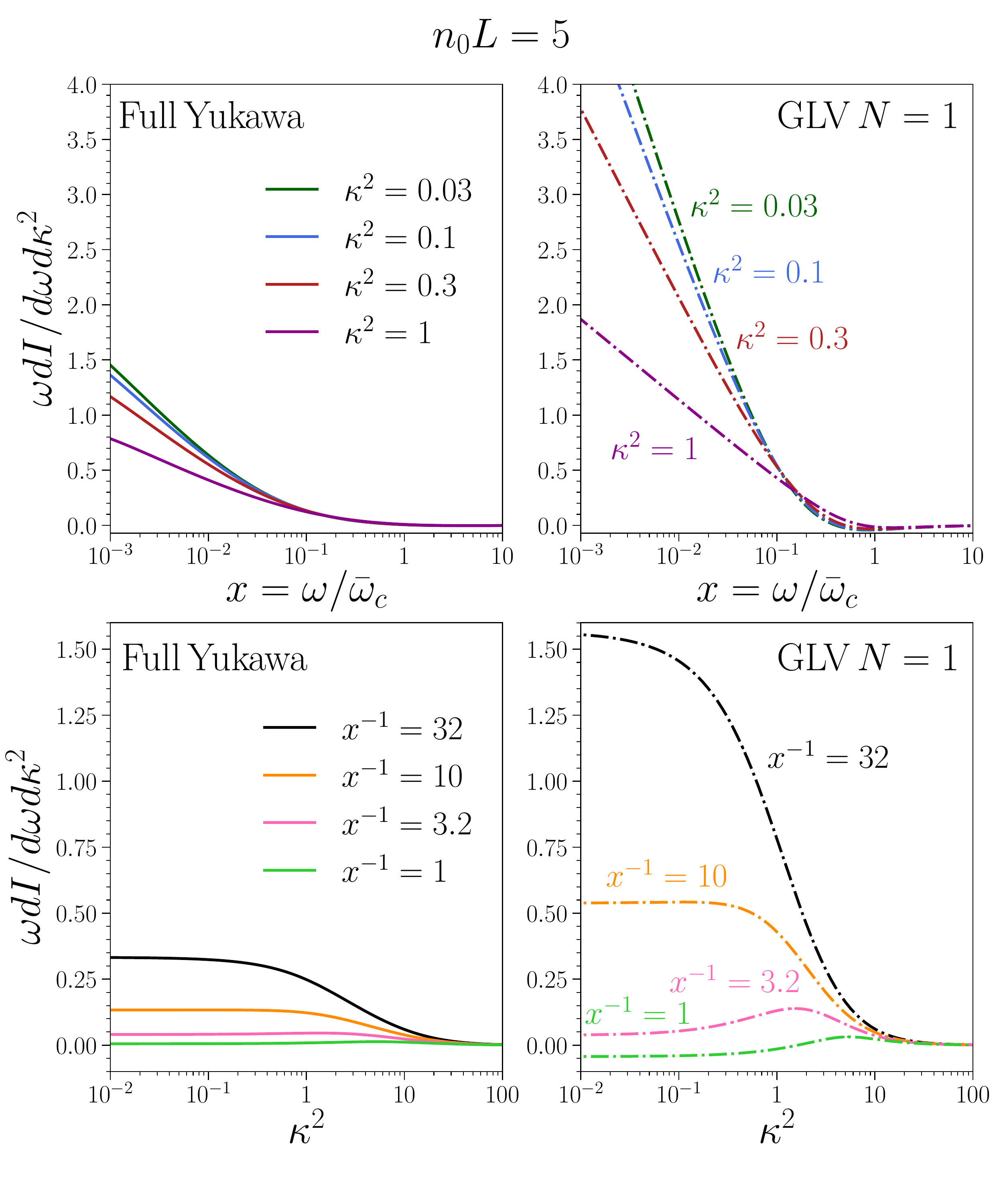}
\vspace*{-5mm}
\caption{Full medium-induced gluon radiation $k{_\perp}$-differential spectrum for a medium with $n_0L=5$ for the Yukawa-type interaction  (left column) compared to  the GLV first opacity approximation (right column). The upper panels show these spectra as function of the
rescaled gluon energy $x=\omega/\bar{\omega_c}$ for fixed values of the rescaled gluon transverse momentum $\kappa = k/\mu$. The lower panels show these spectra versus $\kappa^2$ for fixed values of $x$.}
\label{fig:ktspec_yukawa_glv_n0L5}
\end{figure}

We can now use the above expressions to numerically solve the medium-induced gluon emission spectrum. This collision rate clearly depends only on one  parameter $\mu^2$, but the full evaluation of the spectrum also depends on $n_0$ and $L$. We compute first the full resummed medium-induced $k_{\perp}$-differential spectrum for this interaction. A comparison of our results to the first term in the opacity expansion (GLV $N =1$)\footnote{The GLV spectrum is calculated using the same model for the interaction and therefore depends on the same set of parameters as our approach (see appendix \ref{sec:glv}).} is shown in figures~\ref{fig:ktspec_yukawa_glv_n0L1}~and~\ref{fig:ktspec_yukawa_glv_n0L5}, for two different values of $n_0L=1,5$, respectively. Note that, though the plots are labeled by their value of $n_0L$ only, it has to be kept in mind that the two other parameters involved in the evaluation are implicit in the definition of $\kappa^2 = k^2/\mu^2$ and $x = 2 \omega/(\mu^2L)$.

At large energy and transverse momentum, the magnitude of the full spectrum (left panels of figures~\ref{fig:ktspec_yukawa_glv_n0L1}~and~\ref{fig:ktspec_yukawa_glv_n0L5}) and the GLV approximation (right panels) is the same. Such overlap of the high tails in both figures, is expected since in those kinematical regions ($\omega > \bar\omega_c$ and $\kappa > 1$), where the spectrum is known to be suppressed, the interaction is believed to be dominated by a single hard scattering. As we move towards smaller energies ($\omega < \bar \omega_c$) and/or transverse momenta ($\kappa < 1$), the differences between the two approaches start to become more visible. In particular, in these kinematic regions, the full spectrum is significantly smaller than the GLV approximation, given that the latter does not account for the coherent effect of multiple scatterings, thus confirming that the first order in opacity expansion is not a good approximation when moving away from the large transverse momentum and/or large energy region.  With increasing $n_0L$, the differences between the two approaches are even larger (see figure~\ref{fig:ktspec_yukawa_glv_n0L5}).

\begin{figure}
\vspace*{-5mm}
\centering
\includegraphics[scale=0.38]{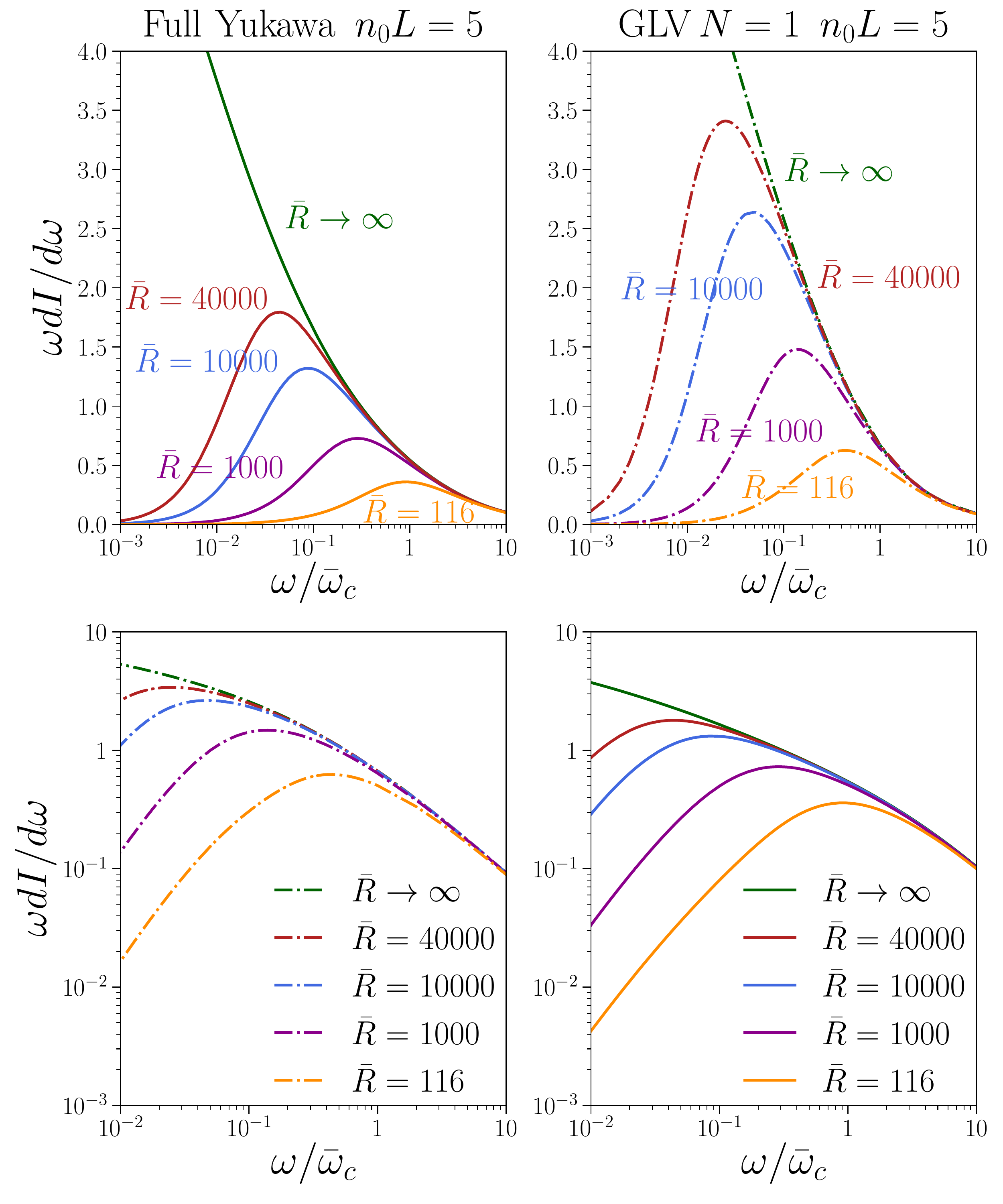}
\vspace*{-2mm}
\caption{Top: full medium-induced gluon energy distribution for the Yukawa collision rate  (left panel) compared to the GLV first opacity approximation (right panel) for different values of $\bar R = \mu^2L^2/2$ for a medium with $n_0L=5$ as a function of  the rescaled gluon energy $\omega/\bar\omega_c$. Bottom: same as top panels in log-log scale.}
\label{fig:spec_yukawa}
\end{figure}

We now present the numerical evaluation of the full resummed medium-induced energy spectrum given by eq.~(\ref{eq:speckinconst}) for the Yukawa collision rate. Even though this energy distribution depends on the same three parameters as the transverse momentum spectrum, we can employ instead the following:
\beq
n_0L \: ,\ \ \ \bar \omega_c= \mu^2L/2 \: ,  \ \  \mathrm{and} \ \ \  \bar{R} = \bar \omega_cL \: ,
\eeq
where the latter can be seen as a dimensionless kinematic constraint on the transverse momentum phase space of the emitted gluon ensuring that $k \leq \omega$. Indeed, the limit $\bar R \rightarrow \infty$ --- which removes this kinematic constraint --- can be viewed as the limit of infinite in-medium path length since it corresponds to $L \rightarrow \infty$ for $\bar \omega_c$ fixed.

In figure \ref{fig:spec_yukawa} we show the comparison of the full resummed medium-induced gluon energy distribution for the Yukawa-type interaction (left panels) with the GLV first opacity (right panels) assuming $n_0L=5$. As previously mentioned, due to the lack of LPM suppression in the GLV approach, for a fixed value of $\bar R$ and $\omega < \bar \omega_c$ the full result is smaller than the first opacity evaluation. In order to illustrate the asymptotic behavior of the two approaches, we show in the bottom panels of  figure~\ref{fig:spec_yukawa} the same spectra using a  logarithmic scale for both axes. From here, it becomes clear that both spectra are suppressed as $\bar \omega_c /\omega$ for $\omega > \bar \omega_c$, but their magnitudes start to differ for $\omega \leq \bar{\omega}_c$.

\begin{figure}[th]
\vspace*{-5mm}
\centering
\includegraphics[scale=0.45]{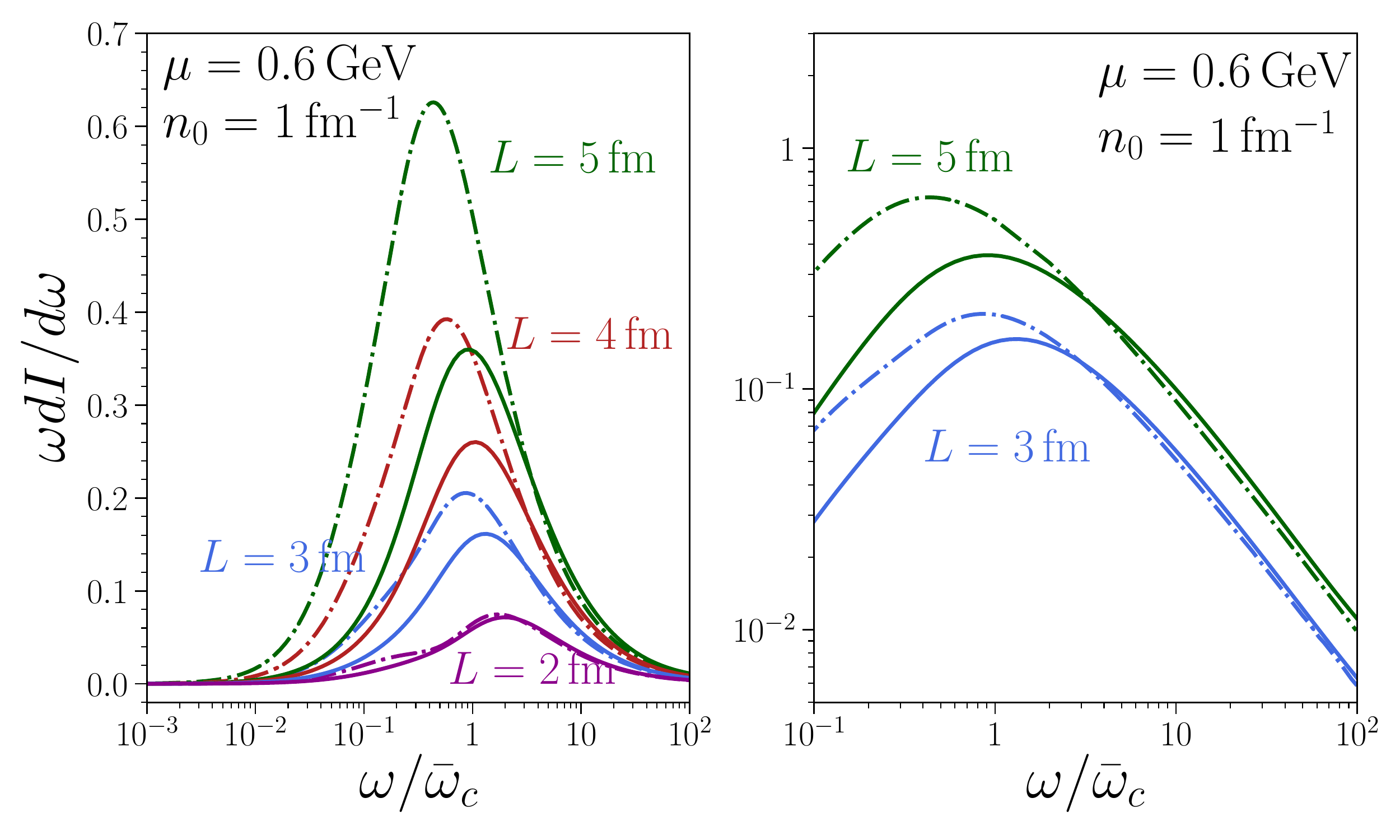}
\vspace*{-2mm}
\caption{Left: full medium-induced gluon energy distribution  for the Yukawa-type interaction (solid lines) compared to the GLV first opacity approximation (dash-dotted lines) with $\mu=0.6$ GeV and linear density $n_0 = 1 \,\mathrm{fm^{-1}}$ for different values of $L$ as a function of the rescaled gluon energy $\omega/\bar\omega_c$. Right: same as left panel for only two values of $L$ in log-log scale.} 
\label{fig:spec_yukawa_severalL}
\end{figure}

A comparison fixing the value of the linear density of scattering centers, $n_0 = 1 \,\mathrm{fm^{-1}}$, and $\mu = 0.6$~GeV while varying the medium path lengths, $L=2,3,4,5$~fm, is shown in figure~\ref{fig:spec_yukawa_severalL} (left). Since the opacity expansion is justified for small values of $n_0L$, for a fixed linear density, the smaller the value of the path length the smaller the discrepancy between the first opacity approximation and the full resummed result. Furthermore, at large gluon energies, the GLV energy distribution is a good approximation of the full resummed solution, since in this kinematical region the process is dominated by a single hard scattering. This can be clearly seen in the right panel of figure~\ref{fig:spec_yukawa_severalL}, where we compare the results for $L=3$~fm and $L=5$~fm using a logarithmic scale for the vertical axis. Again, the differences between the two spectra start to increase significantly for values of $\omega \lesssim \bar\omega_c$.

Now we turn our attention to the comparison between our full resummed results for the Yukawa collision rate and the Gaussian approximation. First, it is important to keep in mind that the correspondence between the parameters used for each evaluation is not straightforward. In principle, the parameter $\hat q$ --- defined through eq.~(\ref{eq:qhat}) --- is directly related to the elastic cross section and can be calculated as its first moment, but the momentum integration involved in this evaluation has a logarithmic divergence which must be regulated by a cut-off. By expanding the dipole cross section in eq.~(\ref{eq:sigmaV_coordinate}), it can be shown that
\beq
\hat{q}L \sim \left(n_0 L \right) \mu^2 \ln\sqrt{\frac{q_{max}}{\mu}}\:,
\label{eq:qhat_approx}
\eeq
where $q_{max}$ is the upper cut-off of the $q$-integral.

Much has been discussed in the literature about the proper choice for this cut-off \cite{Baier:1996sk,Arnold:2009mr,Mehtar-Tani:2019tvy,Mehtar-Tani:2019ygg} sometimes  making it dependent on the gluon energy. Nevertheless, all phenomenological studies which employ the HO approximation regard $\hat q$ as a local property of the medium, independent of the probe and the radiated gluon, and we will do the same here. For illustrative purposes, we fix $\hat{q}L = 1.3 \left(n_0 L \right) \mu^2$. Changing this particular numerical factor does not improve the agreement between the two approaches.

Additionally, the HO approach depends only on two parameters ($\hat q$ and $L$) as opposed to the full evaluation which depends on three ($n_0$, $\mu^2$, and $L$). Different combinations of $n_0$ and $\mu^2$ can correspond to the same value of $\hat q$, as shown in eq.~(\ref{eq:qhat_approx}), which complicates further the choice of parameters for meaningful comparisons. In the following, we use reasonable sets of parameters to produce comparative plots which allow us to highlight differences between both approaches without intending an exhaustive analysis. Specifically, we take $n_0L=5$ and $L=6$ fm, a set of parameters for which the effects of multiple scatterings are expected to be important.
 
\begin{figure}
\vspace*{-5mm}
\centering
\includegraphics[scale=0.45]{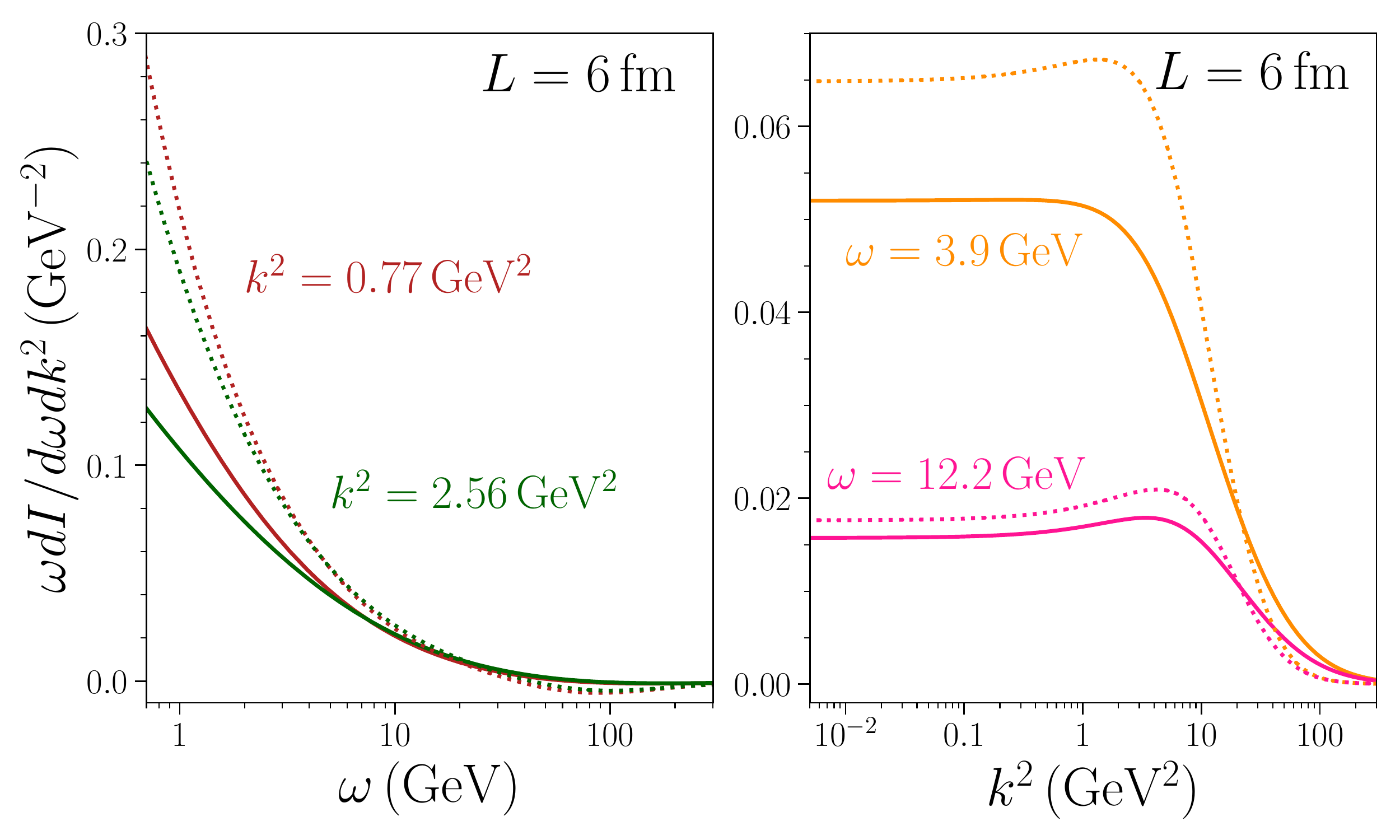}
\vspace*{-2mm}
\caption{Full medium-induced gluon $k_{\perp}$-differential spectrum  for the Yukawa collision rate with $\mu = 1.6\, \mathrm{GeV}$ for a medium of $L= 6$ fm and $n_0L=5$ (solid lines) compared to the evaluation in the harmonic approximation (dotted lines) with $\hat{q} = 2.8 \, \mathrm{GeV^2/fm}$. The left panel shows these two evaluations as function of the emitted gluon energy $\omega$ for two different values of the gluon transverse momenta $k$. The right panel shows these spectra versus $k^2$ for two different values of $\omega$.}
\label{fig:ktspec_yukawa_HO_n0L5}
\end{figure}

The results for the $k_{\perp}$-differential in-medium spectrum for the full resummed solution (solid) and the HO approximation (dash-dotted) are shown in figure~\ref{fig:ktspec_yukawa_HO_n0L5}  with $\mu = 1.6$ GeV (or $\bar\omega_c = 39$ GeV) for the full result and $\hat q = 2.8 \, \mathrm{GeV^2/fm}$ (or $\omega_c \equiv \hat{q}L^2/2= 256$ GeV) for the HO result.\footnote{The formulae used to produce the HO curves can be found e.g. in the appendix of \cite{Salgado:2003gb}.} The left panel shows their evolution with energy at fixed transverse momentum and the right panel their evolution with transverse momentum for two different values of the energy. Even though the full result does not completely agree with the HO approximation, the differences are much smaller than the ones seen in the comparison with the GLV result shown in figure~\ref{fig:ktspec_yukawa_glv_n0L5}. In the left panel of figure~\ref{fig:ktspec_yukawa_HO_n0L5} the spectrum for the HO approximation  seems to diverge faster than the full result for small $\omega$. This is a region where the validity of the formalism is not guaranteed since one of the main assumptions is that the transverse momentum of the gluon must be much smaller than its energy. In addition, in the right panel, it is visible that the HO result falls much faster than the full result at large $k^2$. Such difference is expected since it is well known that the HO approximation does not reproduce the asymptotic behavior at large transverse momenta.

\begin{figure}[th]
\centering
\vspace*{-5mm}
\includegraphics[scale=0.4]{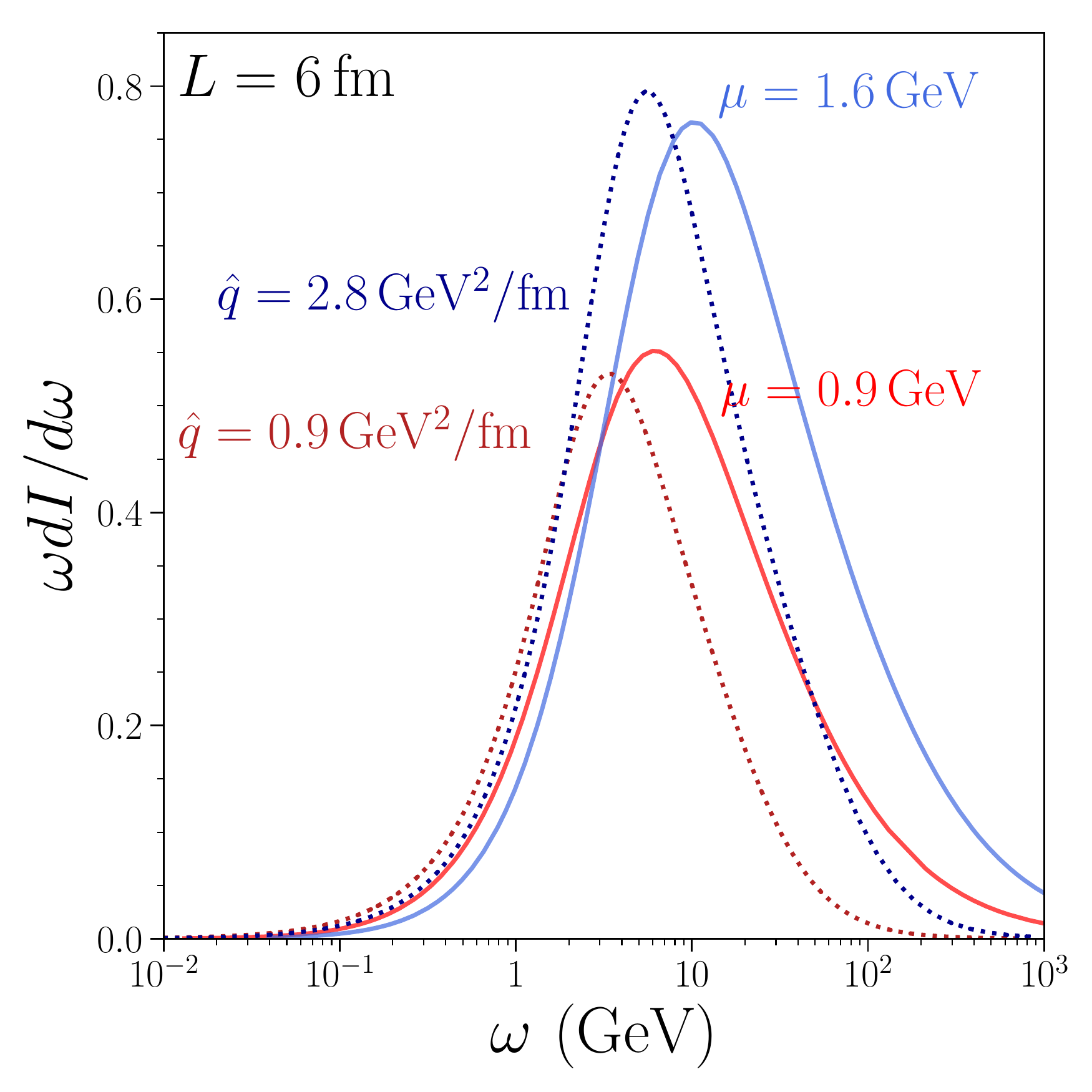}
\vspace*{-2mm}
\caption{Solid lines: full medium-induced gluon energy distribution for the Yukawa-type interaction with $\mu=1.6$ GeV (blue) and $\mu=0.9$ GeV (red) for a medium with $n_0L=5$ and $L=6$ fm as a function of the gluon energy $\omega$. Dotted lines: medium-induced gluon energy distribution in the harmonic approximation for $\hat{q} = 2.8 \,\mathrm{GeV^2/fm}$ (blue) and $\hat{q} = 0.9\, \mathrm{GeV^2/fm}$ (red) and $L=6$ fm versus $\omega$.} 
\label{fig:spec_yukawa_harmonic}
\end{figure}

We now turn to the medium-induced gluon energy distribution in figure~\ref{fig:spec_yukawa_harmonic}. As in the previous figures, the solid lines correspond to the full resummed result for $\mu =1.6$ GeV (or $\bar\omega_c=39$ GeV in blue) and $\mu = 0.9$ GeV (or $\bar\omega_c =12.3$ GeV in red). The dotted lines represent the HO evaluation for the corresponding values of $\hat{q}$ according to $\hat{q}L = 1.3 \left(n_0 L \right) \mu^2$, i.e., $\hat{q} = 2.8 \,\mathrm{GeV^2/fm}$ (blue) and $\hat{q} = 0.9\, \mathrm{GeV^2/fm}$ (red). These yield an $\omega_c \equiv \hat{q}L^2/2 = 256$ GeV (blue) and  $\omega_c= 82$ GeV (red).

For large energies ($\omega > \omega_c$), contrarily to the GLV evaluation, the HO differs significantly from the full resummed result, since this region is dominated by single hard scattering and thus the Gaussian approximation is not well justified. In fact, the HO result decreases proportionally to $1/\omega^2$ in this kinematic region (see for instance \cite{Salgado:2003gb}), in contrast with the $1/\omega$-behavior of the GLV and full resummed evaluations, as it was shown in figure~\ref{fig:spec_yukawa_severalL}. For lower energies ($\omega < \omega_c$), the HO approximation is expected to capture the main physical effects of the emissions process and thus it has been massively used to describe (with relative success so far) the overall behaviour of the gluon emission spectrum at low energies. However, since it is not a formal limit of more general equations, we do not expect a perfect agreement between the Gaussian and the full resummed evaluations in any kinematic region. Although the comparison between our approach and the HO result must be taken with care (as they do not involve the same parameters), we can observe in figure~\ref{fig:spec_yukawa_harmonic} that for $\omega < \omega_c$ the two evaluations differ. It is important to note here that a better agreement between both calculations is found for energies below the peak of the spectra, corresponding to the region where the kinematical constraint $k<\omega$ ($R$ finite) is more important. Furthermore, it is worth noticing that for $\omega < \bar\omega_c $ (with $\bar\omega_c=39$ GeV for the blue curve and  $\bar\omega_c=12.3$ GeV for the red one), the HO result is very close to the full evaluation, while in the same kinematic region the GLV approximation disagrees substantially with the full resummed result (see the right panel of figure~\ref{fig:spec_yukawa_severalL}). 

We have checked that these observations are  independent of our choice of the numerical value of the logarithmic factor used to fix the parameters in eq.~(\ref{eq:qhat_approx}). Changing this factor moves the position of the peak  while keeping the form of the low-$\omega$ tail. The discrepancies we find between our approach and the HO evaluation are in agreement with the results in \cite{Feal:2018sml} where a numerical evaluation through Monte Carlo techniques was shown to significantly differ from the HO approach. 

Our results highlight the shortcomings of the Gaussian approximation and emphasize the necessity of implementing the full resummed spectrum for phenomenological studies. From a theoretical point of view, it would be interesting to thoroughly explore the parameter space to determine the kinematic regions where the multiple soft scattering approximation --- or possible improvements \cite{Mehtar-Tani:2019tvy,Mehtar-Tani:2019ygg,Barata:2020sav} --- provide an accurate description of the emission process. This is beyond the scope of the current manuscript and will be left for future work.

\subsection{Hard thermal loop interaction}

To illustrate the flexibility of our approach, we also implement the collision rate derived from hard thermal loop calculations which should, in principle, provide a more accurate description of the thermal interactions. For this purpose, we take
\beq
\frac{1}{2}n\;V(\vec{q}) = \frac{g_s^2N_c m_D^2T}{\vec{q}^2(\vec{q}^2+m_D^2)}\: ,
\label{eq:HTL}
\eeq
which was obtained at leading-order in the coupling in thermal field theory in a weakly-coupled medium \cite{Aurenche:2002pd}. 

The angular integrations given by eqs.~(\ref{eq:V1})~and~(\ref{eq:V2}) can be performed analytically for this collision rate, giving
\begin{align}
\frac{1}{2}\tilde n\, \tilde V_1(q,p;x) &= g_s^2N_cT\left(\frac{1}{|p^2-q^2|}-\frac{1}{\sqrt{(p^2+q^2+1/x)^2-4p^2q^2}}\right)\:,
\label{eq:V1tildeHTL} \\
\frac{1}{2}\tilde n\, \tilde V_2(q,p;x) &= \frac{g_s^2N_cT}{2pq}\left(\frac{p^2+q^2}{|p^2-q^2|}-\frac{p^2+q^2+1/x}{\sqrt{(p^2+q^2+1/x)^2-4p^2q^2}}\right)\:. 
\label{eq:V2tildeHTL}
\end{align}

One difficulty with this potential is that it is divergent for $p=q$. These divergences always disappear when trying to solve both the initial conditions and the differential equations. Nevertheless, care must be taken to avoid a numerical evaluation at that particular point.

Plugging eqs.~(\ref{eq:V1tildeHTL})~and~(\ref{eq:V2tildeHTL}) in eq.~(\ref{eq:initcondg}) gives
\beq
g_x(s,l;s,p) = \frac{g_s^2N_cTL}{2l^2}\left[\text{sgn}(l-p)+\frac{-l^2+p^2+1/x}{\sqrt{(l^2+p^2+1/x)^2-4l^2p^2}}\right]\: .
\eeq
The result is clearly discontinuous, but it does not induce any singularities in the subsequent steps of the calculations. In practice, the discontinuous term will be set to zero at $p = l$. This result will enter the integrations over $l^2$ from zero to (possibly) infinity, so it is important to note that even though $g_x(s,l;s,p)$ may seem to behave like $1/l^2$ in both of those endpoints, the factor in brackets goes to zero, thus guaranteeing the integration of $g_x(s,l;s,p)$ over $l^2$ to be convergent. 

This initial condition must be evolved with eq.~(\ref{eq:diffeqg}), where $\tilde V_1(l,q;x)$ is singular for $l=q$, but this singularity does not play any role since, again, the term in brackets goes to zero in that limit. The resulting integrand will have a discontinuity at that point and will be assigned the average value between left-handed and right-handed limits.  Similarly to the previous cases, each of the two terms in the right-hand-side of eq.~(\ref{eq:dimlessktdiff}) has a divergence, but these divergences cancel out when the sum of the two terms is considered. Again, there will be a discontinuity which will be handled in a similar manner as the one appearing in eq.~(\ref{eq:diffeqg}).

It is worth noticing that the HTL collision rate given by eq.~(\ref{eq:HTL}) depends on the Debye mass $m_D$ and the medium temperature $T$. By replacing eqs.~(\ref{eq:V1tildeHTL})~and~(\ref{eq:V2tildeHTL}) in the differential equations of section~\ref{subsec:diffeqs}, it is straightforward to see that the full resummed spectrum for this type of interaction depends on the following three free parameters $T$, $m_D^2$, and $L$.\footnote{Note that the Debye mass can be written in terms of the temperature as $m_D^2 = (1+N_f/6)\,g_s^2T^2$ reducing the total number of free parameters from three to two. For convenience, we keep the three independent parameters. In a subsequent paper, where we will apply our results to phenomenology, this relation will be taken into account.} For the energy distribution we will make use instead of
\beq
TL \: , \ \ \ \bar\omega_c^H=m_D^2L/2 \: , \ \ \mathrm{and} \ \ \  \bar R_H =\bar \omega_c^HL\:.
\eeq

\begin{figure}[th]
\centering
\vspace*{-5mm}
\includegraphics[scale=0.38]{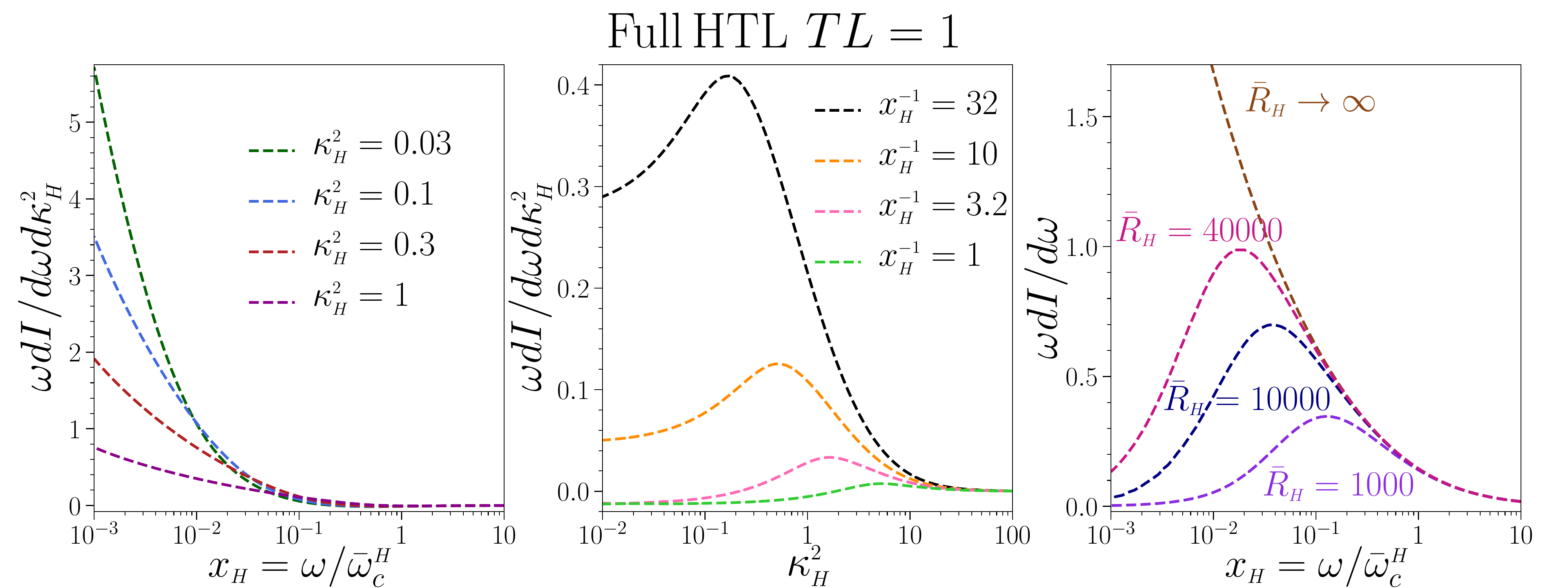}
\vspace*{-3mm}
\caption{Left: full medium-induced $k_{\perp}$-differential gluon radiation  spectrum for the HTL collision rate for a medium with $TL=1$ as a function of rescaled gluon energy $x_H=\omega/\bar\omega_c^H$ for fixed values of the rescaled gluon transverse momentum $\kappa_H = k/m_D$. Center: same as left panel versus $\kappa^2_H$ for fixed values of $x_H$. Right: energy distribution  for the HTL collision rate for a medium with $TL=1$ for different values of $\bar R_H= m_D^2L^2/2$ as a function of $x_H$.}
\label{fig:spec_htl}
\end{figure}

For completeness, we now show the full resummed transverse momentum and energy-dependent in-medium distributions for the HTL collision rate for a medium with $TL=1$ in figure~\ref{fig:spec_htl}. Most of the features previously discussed for the Yukawa potential are also visible in this case, but there are clear differences in the shapes of their $k_T-$differential spectra,  both as a function of energy (left panel) and transverse momentum  (center panel). In order to pursue a more quantitative comparison between the Yukawa and the HTL cases it is necessary to establish a meaningful correspondence between their  respective parameters. The behavior of the collision rate at large transverse momentum is fixed by the Coulomb nature of the interaction at small distances and therefore, by directly comparing eqs.~(\ref{eq:yukawaPotential})~and~(\ref{eq:HTL}), it is clear that one needs to require $n_0\mu^2=\alpha_sN_cTm_D^2$. Since this constraint is not enough to establish a one-to-one correspondence between the sets of parameters of both collisions rates, we decide to follow the suggestion of \cite{Barata:2020sav} where it is shown that one can match 
 their  corresponding dipole cross-sections at leading logarithmic accuracy (see eq.~(\ref{eq:sigmaV_coordinate})) for small dipole sizes by also imposing the condition $m_D^2=e\,\mu^2$. We then have
\beq
TL = n_0L/(e\,\alpha_sN_c)\:,\quad \bar\omega_c^H = e\,\bar\omega_c\:,\quad x_H = x/e\:,\quad \mathrm{and}\quad \kappa_H^2 = \kappa^2/e\:.
\label{eq:matchparam}
\eeq

\begin{figure}
\centering
\vspace*{-5mm}
\includegraphics[scale=0.37]{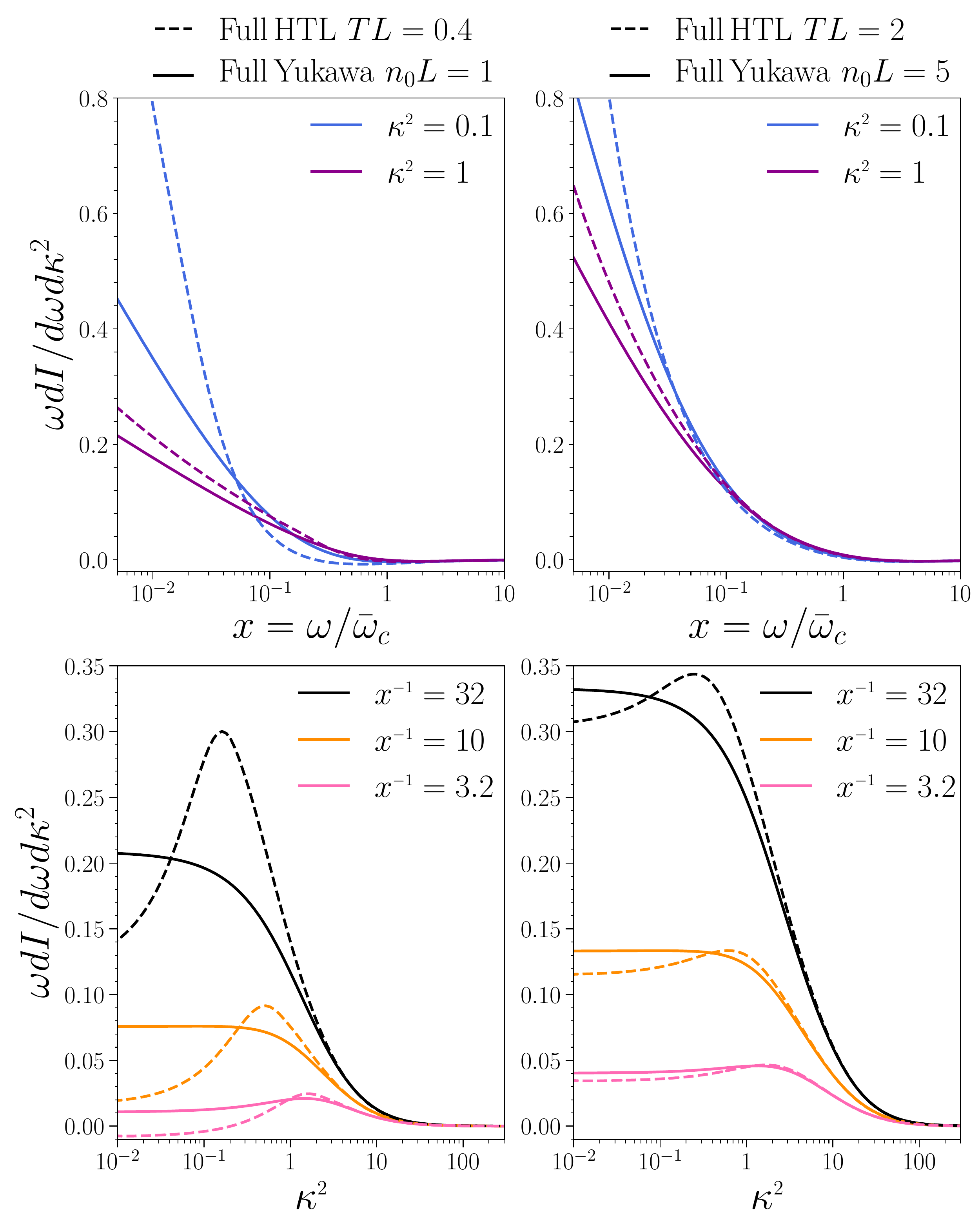}
\vspace*{-1mm}
\caption{Left column: full medium-induced $k_{\perp}$-differential gluon radiation spectrum for the Yukawa-type interaction with $n_0L$= 1 (solid lines) and the HTL interaction with $TL=0.4$ (dashed lines) as a function of $x = \omega/\bar \omega_c=2\omega/\mu^2L$ for different values of $\kappa = k/\mu$ (top) and versus $\kappa^2$ for different values of $x$ (bottom). Right column: same as left panel for the Yukawa collision rate with $n_0L$= 5 (solid lines) and for the HTL collision rate with $TL=2$ (dashed lines).}
\label{fig:htl_vs_yukawa}
\end{figure}

In figure~\ref{fig:htl_vs_yukawa} we show the comparison between the two models for the parton-medium interaction for the $k_\perp$-differential spectrum. The Yukawa case is plotted in solid lines for $n_0L=1$ in the left panels and $n_0L=5$ in the right panels. The HTL case is plotted in dashed lines for the corresponding parameters according to eq.~(\ref{eq:matchparam}), i.e. $TL=0.4$ in the left panels and $TL=2$ in the right panels. Notice that the plots are presented in terms of  Yukawa variables $x$ and $\kappa$, meaning that to produce the curves for the HTL case one has to first rescale these variables according to eq.~(\ref{eq:matchparam}). In the left panels one can see that the agreement between the two collision rates is confined to the high-energy and high-momentum tails, while in the right panels the differences between both evaluations at low and medium energies and transverse momenta substantially decrease. This suggests that when the medium gets larger/denser  the details of the interaction become less important. Finally, we would like to acknowledge that a more exhaustive study of the matching between both interactions would be very interesting and could have a meaningful impact on phenomenological analyses, thus deserving a full separate study.

\section{Conclusions and outlook}
\label{sec:conclusions}

In this work, by using Schwinger-Dyson type equations, we derive an analytical expression for the medium-induced gluon radiation spectrum in the soft limit. Our expression contains the full resummation of multiple scatterings and can be used for any realistic parton-medium interaction without further assumptions, thus providing robust results outside the usually employed multiple soft or single hard scattering approximations. The final outcome is a set of differential equations eqs.~(\ref{eq:dimlessktspec})--(\ref{eq:initcondg}) that can be easily solved and are not as computationally demanding as previous approaches.

In this manuscript, we first use the Yukawa  parton-medium interaction model to compute the full resummed transverse momentum and energy-dependent in-medium gluon emission spectra. We also compare our results with those obtained within the well known HO and GLV approximations, finding the differences among them significant. More specifically, the full resummed spectrum is smaller than the corresponding first opacity evaluation, due to the lack of LPM suppression in the latter approach. However, for large gluon transverse momentum and energy, the GLV limit agrees with the full resummed result, since in this kinematic region the process is dominated by a single hard scattering. The harmonic approximation also differs substantially from the full resummed calculations. Specifically, at high gluon energies and transverse momenta, the harmonic oscillator spectra go much faster to zero than the full results, thus yielding a softer energy distribution.  Nevertheless, at lower gluon energies, the HO evaluation is a better approximation to the full result than the GLV, highlighting the importance of including the effects of multiple scatterings.

In order to show the flexibility of our approach, we also compute the full resummed spectra for the hard thermal loop parton-medium interaction model. Furthermore, we compare the HTL and Yukawa transverse momentum spectra by using a (leading order) map between their respective set of parameters which ensures that the high energy and transverse momenta tails of the spectra for both interaction models coincide. Our results seem to suggest that the differences found between both evaluations decrease significantly for larger/denser media.

The method of evaluation presented here has a great potential to improve the study of jet quenching observables. In future works we plan to expand the reach of our calculations in two main directions: relaxing approximations that will allow us to improve current phenomenological tools, and adapt our formalism to allow precise numerical evaluations of calculations sensitive to the effect of multiple scatterings with the medium.

First, we will relax the soft gluon approximation, thus allowing emitted gluons to take a finite fraction of the energy, then we will explore the case of non-static media. With these two improvements in hand, the formalism can then be used to calculate distributions of energy loss under realistic conditions, which can be used, for instance, to compute the nuclear modification factor and the high transverse momentum azimuthal anisotropies.

On the other hand, we will apply our formalism to evaluate the radiation pattern of an antenna, as a first step towards improving the precision of the evaluations aimed at understanding the role of color coherence in the description of jet and intra-jet observables.

\acknowledgments
We thank N\'estor Armesto, Xabier Feal, Carlos Salgado, and Ricardo V\'azquez for helpful discussions.  CA was supported by the US Department of Energy contract DE-AC05-06OR23177, under which Jefferson Science Associates, LLC operates Jefferson Lab. LA was supported by Funda\c{c}\~{a}o para a Ci\^{e}ncia e Tecnologia (FCT - Portugal) under project DL57/2016/CP1345/CT0004 and CERN/FIS-PAR/0022/2017.

\appendix
\section{Details of the derivation in section~\ref{subsec:reorganization}}
\label{sec:derivation}

We outline here the steps needed to show how eq.~(\ref{eq:fullspect}) can be derived from eq.~(\ref{eq:bdmps}). Replacing the Schwinger-Dyson equations satisfied by the kernel and the broadening factor --- eqs.~(\ref{eq:PSD1}) and~(\ref{eq:KSD1}) --- in eq.~(\ref{eq:bdmps}) one gets
\begin{align}
\omega\frac{dI}{d\omega d^2\vec{k}} &= \frac{2\alpha_s C_R}{(2\pi)^2\omega^2} \Re \int_0^\infty dt \, \int_t^\infty dt'\,  \left[\vec{k}^2\, e^{-i \frac{k^2}{2\omega}(t'-t)}  \right. \nonumber\\ 
&\quad-\,\frac{1}{2}  \int_{t}^{t'} ds \,n(s)  \, \int_{\vec{p}\vec{k}_1} \,\vec{p} \cdot \vec{k} \,e^{-i \frac{k^2}{2\omega}(t'-s)} \sigma(\vec{k}-\vec{k}_1)\widetilde{\cK}(s,\vec{k}_1;t,\vec{p})  \nonumber\\
&\quad-\,\frac{1}{2}  \int_{t'}^{\infty} ds \,n(s) \, \int_{\vec{p}\vec{k}_1} \,\vec{p}^2 \,e^{-i \frac{p^2}{2\omega}(t'-t)} \sigma(\vec{k}_1-\vec{p}) \cP(\infty,\vec{k};s,\vec{k}_1) \nonumber\\ 
&\quad +\,\frac{1}{4} \int_{t}^{t'} ds_1 \,n(s_1) \, \int_{t'}^{\infty} ds_2 \,n(s_2) \, \int_{\vec{p}\vec{q}\vec{k}_1\vec{k}_2}\,\vec{p} \cdot \vec{q}  \; e^{-i \frac{q^2}{2\omega}(t'-s_1)}\nonumber \\ 
&\left.\qquad\times\, \sigma(\vec{q}-\vec{k}_1) \, \sigma(\vec{k}_2-\vec{q}) \,\widetilde{\cK}(s_1,\vec{k}_1;t,\vec{p}) \, \cP(\infty,\vec{k};s_2,\vec{k}_2)\right] \:. \label{eq:firstexp}
\end{align}

First of all, we discard the first term in eq.~(\ref{eq:firstexp}) since it is the vacuum contribution and we only want to keep track of the medium-induced radiation. In all the remaining terms, the $t'$-dependence of the integrand is given by just a phase factor, so we change the order of integration to perform this integral first.

With the new order of the integrals, the $t'$-integral in the second term --- second line of eq.~(\ref{eq:firstexp}) ---, goes from $s$ to $\infty$. It is important to recall now that a regularization procedure to avoid divergences at late times, in the form of a factor $e^{-\epsilon t'}$, has been omitted in eq.~(\ref{eq:bdmps}). When this factor is properly taken into account, the evaluation of the $t'$-integral of this term in the upper limit vanishes. This is equivalent to include the appropriate $i\epsilon$ prescription in the free propagators. The contribution of the second term of eq.~(\ref{eq:firstexp}) after performing $t'$-integral is then given by
\beq
\frac{2\alpha_s C_R}{(2\pi)^2\omega}\Re \int_0^\infty dt \, \int_{t}^{\infty} ds \,n(s)  \int_{\vec{p}\vec{k}_1} \,i\frac{\vec{p}\cdot\vec{k}}{\vec{k}^2}\sigma(\vec{k}-\vec{k}_1)\,\widetilde{\cK}(s,\vec{k}_1;t,\vec{p})\: .
\label{eq:apendix2}
\eeq

For the third term in eq.~(\ref{eq:firstexp}), the integration limits of the $t'$-integral are finite (from $t$ to $s$), hence, the integration is straightforward. It is important to notice  that, since $\cP$ is always real, the evaluation of this integral at the lower limit yields a purely imaginary term, thus not contributing to the final result. 

The $t'$-integral of the last term in eq.~(\ref{eq:firstexp}) is trivial and both terms survive. The final result of integrating in $t'$ eq.~(\ref{eq:firstexp}), excluding the vacuum contribution, is given by
\begin{align}\label{eq:A2}
\omega\frac{dI}{d\omega d^2\vec{k}} &= \frac{2\alpha_s C_R}{(2\pi)^2\omega} \Re \int_0^\infty dt \,\left[ \int_{t}^{\infty} ds \,n(s) \int_{\vec{p}\vec{k}_1} \,i\frac{\vec{p}\cdot\vec{k}}{\vec{k}^2}\sigma(\vec{k}-\vec{k}_1)\,\widetilde{\cK}(s,\vec{k}_1;t,\vec{p}) \right. \nonumber\\
&-  \int_{t}^{\infty} ds \,n(s) \, \int_{\vec{p}\vec{k}_1} i\,e^{-i \frac{p^2}{2\omega}(s-t)} \sigma(\vec{k}_1-\vec{p}) \cP(\infty,\vec{k};s,\vec{k}_1) \nonumber\\ 
& +\,\frac{1}{2} \int_{t}^{\infty} ds_1 \,n(s_1) \, \int_{s_1}^{\infty} ds_2 \,n(s_2) \, \int_{\vec{p}\vec{q}\vec{k}_1\vec{k}_2}i\frac{\vec{p} \cdot \vec{q}}{\vec{q}^2}  \;\left( e^{-i \frac{q^2}{2\omega}(s_2-s_1)}-1\right)\nonumber \\ 
&\left.\times \sigma(\vec{q}-\vec{k}_1) \, \sigma(\vec{k}_2-\vec{q}) \,\widetilde{\cK}(s_1,\vec{k}_1;t,\vec{p}) \, \cP(\infty,\vec{k};s_2,\vec{k}_2)\right] \:.
\end{align}

To finalize, the terms with and without phases can be recombined as follows: the contribution without phase in the third term  with the first term by using eq.~(\ref{eq:PSD1}); and the contribution which do contains a phase in the third term with the second term by using eq.~(\ref{eq:KSD1}). With such modifications, eq.~(\ref{eq:A2}) yields the following compact result
\begin{align}
\omega\frac{dI}{d\omega d^2\vec{k}} = \frac{2\alpha_s C_R}{(2\pi)^2\omega} \Re\int_0^\infty dt&\int_t^\infty ds\; n(s) \int_{\vec{p}\vec{q}\vec{l}}i\vec{p}\cdot\left(\frac{\vec{l}}{\vec{l}^2}-\frac{\vec{q}}{\vec{q}^2}\right)\sigma(\vec{l}-\vec{q})\widetilde\cK(s,\vec{q};t,\vec{p})\cP(\infty,\vec{k};s,\vec{l})
\:.
\end{align}

Changing the order of integration in $t$ and $s$ we arrive at eq.~(\ref{eq:fullspect}).

\section{Towards GLV limit}
\label{sec:glv}

In this appendix we show how to get the GLV limit from our expressions for the full resummed spectrum. To get the single scattering contribution one can take as starting point eq.~\eqref{eq:fullspect} and replace $\cP$ and $\widetilde\cK$ with their vacuum versions. This is equivalent to taking eq.~\eqref{eq:dimlessktspec} but not solving the differential eqs.~\eqref{eq:dimlessktdiff}~and~\eqref{eq:diffeqg}, and instead using only the initial conditions. Following this approach we get for the GLV spectrum
\beq
x\frac{dI}{dx d\kappa^2} = \frac{\alpha_s C_R}{2\pi^2}\frac{L}{x}\Re\int_0^1 ds\;\tilde n(s)\int_0^s dt\int_0^\infty dp\;ip\,e^{-ip^2(s-t)} \left[\tilde V_1(\kappa/\sqrt{x},p;x) - \frac{\sqrt{x}\,p}{\kappa}\tilde V_2(\kappa/\sqrt{x},p;x)\right]\; ,
\eeq
with $\tilde V_i$'s given by eqs.~\eqref{eq:V1}~and~\eqref{eq:V2}. For the case of a uniform medium, as considered in section~\ref{sec:results}, the integrals over $t$ and $s$ can be easily performed, giving
\beq
x\frac{dI}{dx d\kappa^2} = \frac{\alpha_s C_R}{2\pi^2}\frac{n_0L}{x}\int_0^\infty dp\;\frac{p^2-\sin p^2}{p^3}\left[\tilde V_1(\kappa/\sqrt{x},p;x) - \frac{\sqrt{x}\,p}{\kappa}\tilde V_2(\kappa/\sqrt{x},p;x)\right]\; .
\eeq
Using the Yukawa-type interaction, we get
\beq
x\frac{dI}{dx d\kappa^2} = \frac{4\alpha_s C_R}{\pi}n_0L\int_0^\infty dp\;\frac{p^2-\sin p^2}{p^3}\frac{\kappa^2 - xp^2 +1}{[(\kappa^2 + xp^2+1)^2-4x\kappa^2p^2]^{3/2}}\; .
\eeq
This is the formula used for the curves in the right panels of figures~\ref{fig:ktspec_yukawa_glv_n0L1}~and~\ref{fig:ktspec_yukawa_glv_n0L5}. To calculate the energy spectrum, we can integrate over the transverse momentum with the constraint $\kappa^2<\bar Rx^2/2$ to get
\beq
x\frac{dI}{dx} = \frac{4\alpha_s C_R}{\pi}n_0L\int_0^\infty dp\;\frac{p^2-\sin p^2}{p^3}\left[\frac{1}{xp^2+1}-\frac{1}{\sqrt{\left(\frac{1}{2}\bar Rx^2+xp^2+1\right)^2-2x^3\bar Rp^2}}\right]\; .
\eeq
This result allow us to generate the curves shown in the right panels of figures~\ref{fig:spec_yukawa}~and~\ref{fig:spec_yukawa_severalL}.

\providecommand{\href}[2]{#2}\begingroup\raggedright\endgroup


\begin{thebibliography}{10}

\bibitem{Apolinario:2017sob}
L.~Apolinário, J.~G. Milhano, G.~P. Salam and C.~A. Salgado, \emph{{Probing
  the time structure of the quark-gluon plasma with top quarks}},
  \href{https://doi.org/10.1103/PhysRevLett.120.232301}{\emph{Phys. Rev. Lett.}
  {\bfseries 120} (2018) 232301}
  [\href{https://arxiv.org/abs/1711.03105}{{\ttfamily 1711.03105}}].

\bibitem{Andres:2019eus}
C.~Andres, N.~Armesto, H.~Niemi, R.~Paatelainen and C.~A. Salgado, \emph{{Jet
  quenching as a probe of the initial stages in heavy-ion collisions}},
  [\href{https://arxiv.org/abs/1902.03231}{{\ttfamily 1902.03231}}].

\bibitem{Landau:1953um}
L.~D. Landau and I.~Pomeranchuk, \emph{{Limits of applicability of the theory
  of bremsstrahlung electrons and pair production at high-energies}},
  {\emph{Dokl. Akad. Nauk Ser. Fiz.} {\bfseries 92} (1953) 535}.

\bibitem{Migdal:1956tc}
A.~B. Migdal, \emph{{Bremsstrahlung and pair production in condensed media at
  high-energies}}, \href{https://doi.org/10.1103/PhysRev.103.1811}{\emph{Phys.
  Rev.} {\bfseries 103} (1956) 1811}.

\bibitem{MehtarTani:2010ma}
Y.~Mehtar-Tani, C.~A. Salgado and K.~Tywoniuk, \emph{{Anti-angular ordering of
  gluon radiation in QCD media}},
  \href{https://doi.org/10.1103/PhysRevLett.106.122002}{\emph{Phys. Rev. Lett.}
  {\bfseries 106} (2011) 122002}
  [\href{https://arxiv.org/abs/1009.2965}{{\ttfamily 1009.2965}}].

\bibitem{MehtarTani:2011tz}
Y.~Mehtar-Tani, C.~A. Salgado and K.~Tywoniuk, \emph{{Jets in QCD Media: From
  Color Coherence to Decoherence}},
  \href{https://doi.org/10.1016/j.physletb.2011.12.042}{\emph{Phys. Lett.}
  {\bfseries B707} (2012) 156}
  [\href{https://arxiv.org/abs/1102.4317}{{\ttfamily 1102.4317}}].

\bibitem{MehtarTani:2012cy}
Y.~Mehtar-Tani, C.~A. Salgado and K.~Tywoniuk, \emph{{The Radiation pattern of
  a QCD antenna in a dense medium}},
  \href{https://doi.org/10.1007/JHEP10(2012)197}{\emph{JHEP} {\bfseries 10}
  (2012) 197} [\href{https://arxiv.org/abs/1205.5739}{{\ttfamily 1205.5739}}].

\bibitem{Baier:1996kr}
R.~Baier, Y.~L. Dokshitzer, A.~H. Mueller, S.~Peigne and D.~Schiff,
  \emph{{Radiative energy loss of high-energy quarks and gluons in a finite
  volume quark - gluon plasma}},
  \href{https://doi.org/10.1016/S0550-3213(96)00553-6}{\emph{Nucl. Phys.}
  {\bfseries B483} (1997) 291}
  [\href{https://arxiv.org/abs/hep-ph/9607355}{{\ttfamily hep-ph/9607355}}].

\bibitem{Baier:1996sk}
R.~Baier, Y.~L. Dokshitzer, A.~H. Mueller, S.~Peigne and D.~Schiff,
  \emph{{Radiative energy loss and p(T) broadening of high-energy partons in
  nuclei}}, \href{https://doi.org/10.1016/S0550-3213(96)00581-0}{\emph{Nucl.
  Phys.} {\bfseries B484} (1997) 265}
  [\href{https://arxiv.org/abs/hep-ph/9608322}{{\ttfamily hep-ph/9608322}}].

\bibitem{Zakharov:1996fv}
B.~G. Zakharov, \emph{{Fully quantum treatment of the Landau-Pomeranchuk-Migdal
  effect in QED and QCD}}, \href{https://doi.org/10.1134/1.567126}{\emph{JETP
  Lett.} {\bfseries 63} (1996) 952}
  [\href{https://arxiv.org/abs/hep-ph/9607440}{{\ttfamily hep-ph/9607440}}].

\bibitem{Zakharov:1997uu}
B.~G. Zakharov, \emph{{Radiative energy loss of high-energy quarks in finite
  size nuclear matter and quark - gluon plasma}},
  \href{https://doi.org/10.1134/1.567389}{\emph{JETP Lett.} {\bfseries 65}
  (1997) 615} [\href{https://arxiv.org/abs/hep-ph/9704255}{{\ttfamily
  hep-ph/9704255}}].

\bibitem{Wiedemann:2000za}
U.~A. Wiedemann, \emph{{Gluon radiation off hard quarks in a nuclear
  environment: Opacity expansion}},
  \href{https://doi.org/10.1016/S0550-3213(00)00457-0}{\emph{Nucl. Phys.}
  {\bfseries B588} (2000) 303}
  [\href{https://arxiv.org/abs/hep-ph/0005129}{{\ttfamily hep-ph/0005129}}].

\bibitem{Feal:2018sml}
X.~Feal and R.~Vazquez, \emph{{Intensity of gluon bremsstrahlung in a finite
  plasma}}, \href{https://doi.org/10.1103/PhysRevD.98.074029}{\emph{Phys. Rev.}
  {\bfseries D98} (2018) 074029}
  [\href{https://arxiv.org/abs/1811.01591}{{\ttfamily 1811.01591}}].

\bibitem{CaronHuot:2010bp}
S.~Caron-Huot and C.~Gale, \emph{{Finite-size effects on the radiative energy
  loss of a fast parton in hot and dense strongly interacting matter}},
  \href{https://doi.org/10.1103/PhysRevC.82.064902}{\emph{Phys. Rev.}
  {\bfseries C82} (2010) 064902}
  [\href{https://arxiv.org/abs/1006.2379}{{\ttfamily 1006.2379}}].

\bibitem{Gyulassy:2000er}
M.~Gyulassy, P.~Levai and I.~Vitev, \emph{{Reaction operator approach to
  nonAbelian energy loss}},
  \href{https://doi.org/10.1016/S0550-3213(00)00652-0}{\emph{Nucl. Phys.}
  {\bfseries B594} (2001) 371}
  [\href{https://arxiv.org/abs/nucl-th/0006010}{{\ttfamily nucl-th/0006010}}].

\bibitem{Arnold:2001ms}
P.~B. Arnold, G.~D. Moore and L.~G. Yaffe, \emph{{Photon emission from quark
  gluon plasma: Complete leading order results}},
  \href{https://doi.org/10.1088/1126-6708/2001/12/009}{\emph{JHEP} {\bfseries
  12} (2001) 009} [\href{https://arxiv.org/abs/hep-ph/0111107}{{\ttfamily
  hep-ph/0111107}}].

\bibitem{CasalderreySolana:2007pr}
J.~Casalderrey-Solana and C.~A. Salgado, \emph{{Introductory lectures on jet
  quenching in heavy ion collisions}}, {\emph{Acta Phys. Polon.} {\bfseries
  B38} (2007) 3731} [\href{https://arxiv.org/abs/0712.3443}{{\ttfamily
  0712.3443}}].

\bibitem{Blaizot:2012fh}
J.-P. Blaizot, F.~Dominguez, E.~Iancu and Y.~Mehtar-Tani, \emph{{Medium-induced
  gluon branching}}, \href{https://doi.org/10.1007/JHEP01(2013)143}{\emph{JHEP}
  {\bfseries 01} (2013) 143} [\href{https://arxiv.org/abs/1209.4585}{{\ttfamily
  1209.4585}}].

\bibitem{Apolinario:2014csa}
L.~Apolinário, N.~Armesto, J.~G. Milhano and C.~A. Salgado,
  \emph{{Medium-induced gluon radiation and colour decoherence beyond the soft
  approximation}}, \href{https://doi.org/10.1007/JHEP02(2015)119}{\emph{JHEP}
  {\bfseries 02} (2015) 119} [\href{https://arxiv.org/abs/1407.0599}{{\ttfamily
  1407.0599}}].

\bibitem{Apolinario:2012vy}
L.~Apolinario, N.~Armesto and C.~A. Salgado, \emph{{Medium-induced emissions of
  hard gluons}},
  \href{https://doi.org/10.1016/j.physletb.2012.10.040}{\emph{Phys. Lett.}
  {\bfseries B718} (2012) 160}
  [\href{https://arxiv.org/abs/1204.2929}{{\ttfamily 1204.2929}}].

\bibitem{Salgado:2002cd}
C.~A. Salgado and U.~A. Wiedemann, \emph{{A Dynamical scaling law for jet
  tomography}},
  \href{https://doi.org/10.1103/PhysRevLett.89.092303}{\emph{Phys. Rev. Lett.}
  {\bfseries 89} (2002) 092303}
  [\href{https://arxiv.org/abs/hep-ph/0204221}{{\ttfamily hep-ph/0204221}}].

\bibitem{Arnold:2009mr}
P.~B. Arnold, \emph{{High-energy gluon bremsstrahlung in a finite medium:
  harmonic oscillator versus single scattering approximation}},
  \href{https://doi.org/10.1103/PhysRevD.80.025004}{\emph{Phys. Rev.}
  {\bfseries D80} (2009) 025004}
  [\href{https://arxiv.org/abs/0903.1081}{{\ttfamily 0903.1081}}].

\bibitem{Salgado:2003gb}
C.~A. Salgado and U.~A. Wiedemann, \emph{{Calculating quenching weights}},
  \href{https://doi.org/10.1103/PhysRevD.68.014008}{\emph{Phys. Rev.}
  {\bfseries D68} (2003) 014008}
  [\href{https://arxiv.org/abs/hep-ph/0302184}{{\ttfamily hep-ph/0302184}}].

\bibitem{Andres:2016iys}
C.~Andrés, N.~Armesto, M.~Luzum, C.~A. Salgado and P.~Zurita, \emph{{Energy
  versus centrality dependence of the jet quenching parameter $\hat{q}$ at RHIC
  and LHC: a new puzzle?}},
  \href{https://doi.org/10.1140/epjc/s10052-016-4320-5}{\emph{Eur. Phys. J.}
  {\bfseries C76} (2016) 475}
  [\href{https://arxiv.org/abs/1606.04837}{{\ttfamily 1606.04837}}].

\bibitem{Burke:2013yra}
{\scshape JET} collaboration, \emph{{Extracting the jet transport coefficient
  from jet quenching in high-energy heavy-ion collisions}},
  \href{https://doi.org/10.1103/PhysRevC.90.014909}{\emph{Phys. Rev.}
  {\bfseries C90} (2014) 014909}
  [\href{https://arxiv.org/abs/1312.5003}{{\ttfamily 1312.5003}}].

\bibitem{Feal:2019xfl}
X.~Feal, C.~A. Salgado and R.~A. Vazquez, \emph{{Jet quenching tests of the QCD
  Equation of State}},  [\href{https://arxiv.org/abs/1911.01309}{{\ttfamily
  1911.01309}}].

\bibitem{Aurenche:2002pd}
P.~Aurenche, F.~Gelis and H.~Zaraket, \emph{{A Simple sum rule for the thermal
  gluon spectral function and applications}},
  \href{https://doi.org/10.1088/1126-6708/2002/05/043}{\emph{JHEP} {\bfseries
  05} (2002) 043} [\href{https://arxiv.org/abs/hep-ph/0204146}{{\ttfamily
  hep-ph/0204146}}].

\bibitem{Moore:2019lgw}
G.~D. Moore and N.~Schlusser, \emph{{Transverse momentum broadening from the
  lattice}}, \href{https://doi.org/10.1103/PhysRevD.101.014505}{\emph{Phys.
  Rev.} {\bfseries D101} (2020) 014505}
  [\href{https://arxiv.org/abs/1911.13127}{{\ttfamily 1911.13127}}].

\bibitem{Mehtar-Tani:2019tvy}
Y.~Mehtar-Tani, \emph{{Gluon bremsstrahlung in finite media beyond multiple
  soft scattering approximation}},
  \href{https://doi.org/10.1007/JHEP07(2019)057}{\emph{JHEP} {\bfseries 07}
  (2019) 057} [\href{https://arxiv.org/abs/1903.00506}{{\ttfamily
  1903.00506}}].

\bibitem{Mehtar-Tani:2019ygg}
Y.~Mehtar-Tani and K.~Tywoniuk, \emph{{Improved opacity expansion for
  medium-induced parton splitting}},
 [\href{https://arxiv.org/abs/1910.02032}{{\ttfamily 1910.02032}}].

\bibitem{Barata:2020sav}
J.~Barata and Y.~Mehtar-Tani,
\emph{{Improved opacity expansion at NNLO for medium induced gluon radiation}},
[\href{https://arxiv.org/abs/2004.02323}{{\ttfamily 2004.02323}}].




\end{thebibliography}
\end{document}